\documentclass[12pt, draftclsnofoot, onecolumn]{IEEEtran}
\usepackage{cite,graphicx,amsmath,amssymb}
\usepackage{subfigure}
\usepackage{citesort}
\usepackage{fancyhdr}
\usepackage{mdwmath}
\usepackage{mdwtab}
\usepackage{balance}
\usepackage{xcolor}
\usepackage{bm}
\usepackage{cite,graphicx,amsmath,amssymb}
\usepackage{subfigure}
\usepackage{citesort}
\usepackage{fancyhdr}
\usepackage{mdwmath}
\usepackage{mdwtab}
\usepackage{balance}
\usepackage{xcolor}
\usepackage{bm}
\usepackage{amssymb}
\usepackage{amsmath}
\usepackage{cite}
\usepackage{url}
\usepackage{xcolor}
\usepackage{cite,graphicx,amsmath,amssymb}
\usepackage{subfigure}
\usepackage{citesort}
\usepackage{fancyhdr}
\usepackage{mdwmath}
\usepackage{mdwtab}
\usepackage{float}
\usepackage{caption}
\usepackage{amsthm}
\usepackage{graphicx}
\usepackage{algorithm}
\usepackage{algorithmic}
\usepackage{multirow}
\usepackage{flafter}
\usepackage{graphicx}
\usepackage{subfigure}
\usepackage{algorithm}
\usepackage{setspace}
\usepackage{algorithmic}
\usepackage{caption}
\usepackage{amssymb}
\usepackage{amsmath}
\usepackage{cite}
\usepackage{url}
\usepackage{xcolor}
\usepackage{cite,graphicx,amsmath,amssymb}
\usepackage{subfigure}
\usepackage{citesort}
\usepackage{fancyhdr}
\usepackage{mdwmath}
\usepackage{mdwtab}
\usepackage{caption}
\usepackage{amsthm}
\usepackage{algorithm}
\usepackage{algorithmic}

\usepackage{multirow}
\usepackage{flafter}
\usepackage{graphicx}
\usepackage{subfigure}
\usepackage{algorithm}
\usepackage{setspace}
\usepackage{algorithmic}
\usepackage{caption}
\newtheorem{proposition}{Proposition}
\newtheorem{remark}{Remark}

\usepackage{amsmath,amsthm,amssymb}

\newtheorem{theorem}{Theorem}

\begin{document}

\title{RIS Enhanced Massive Non-orthogonal Multiple Access Networks: Deployment and Passive Beamforming Design}

\author{
Xiao~Liu,~\IEEEmembership{Student Member,~IEEE,}
Yuanwei~Liu,~\IEEEmembership{Senior Member,~IEEE,}\\
Yue~Chen,~\IEEEmembership{Senior Member,~IEEE,}
and H. Vincent~Poor,~\IEEEmembership{Fellow,~IEEE}

\thanks{

X. Liu, Y. Liu, and Y. Chen are with the School of Electronic Engineering and Computer Science, Queen Mary University of London. (email: x.liu@qmul.ac.uk; yuanwei.liu@qmul.ac.uk; yue.chen@qmul.ac.uk)

H. Vincent Poor is with Department of Electrical Engineering, Princeton University. (email: poor@princeton.edu)

}
}

\maketitle

\vspace{-2cm}

\begin{abstract}
A novel framework is proposed for the deployment and passive beamforming design of a reconfigurable intelligent surface (RIS) with the aid of non-orthogonal multiple access (NOMA) technology. The problem of joint deployment, phase shift design, as well as power allocation is formulated for maximizing the energy efficiency with considering users' particular data requirements. To tackle this pertinent problem, machine learning approaches are adopted in two steps. Firstly, a novel long short-term memory (LSTM) based echo state network (ESN) algorithm is proposed to predict users' tele-traffic demand by leveraging a real dataset. Secondly, a decaying double deep Q-network (${{\text{D}}^{\text{3}}}{\text{QN}}$) based position-acquisition and phase-control algorithm is proposed to solve the joint problem of deployment and design of the RIS. In the proposed algorithm, the base station, which controls the RIS by a controller, acts as an agent. The agent periodically observes the state of the RIS-enhanced system for attaining the optimal deployment and design policies of the RIS by learning from its mistakes and the feedback of users. Additionally, it is proved that the proposed ${{\text{D}}^{\text{3}}}{\text{QN}}$ based deployment and design algorithm is capable of converging within mild conditions. Simulation results are provided for illustrating that the proposed LSTM-based ESN algorithm is capable of striking a tradeoff between the prediction accuracy and computational complexity. Finally, it is demonstrated that the proposed ${{\text{D}}^{\text{3}}}{\text{QN}}$ based algorithm outperforms the benchmarks, while the NOMA-enhanced RIS system is capable of achieving higher energy efficiency than orthogonal multiple access (OMA) enabled RIS system.
\end{abstract}
\vspace{-0.5cm}

\vspace{-0.5cm}

\section{Introduction}

Owing to their capability of proactively modifying the wireless communication environment, reconfigurable intelligent surfaces (RISs), also named as reconfigurable reflect-arrays, large intelligent surfaces (LISs)~\cite{hou2019mimo} or intelligent reflecting surface (IRS), have become a focal point in the wireless communications research field for mitigating a wide range of challenges encountered in diverse wireless networks~\cite{yang2019intelligent}. The RIS is made of electromagnetic material, which can be installed on key points, such as building facades, highway polls, advertising panels, vehicle windows, and even pedestrians' clothes due to the characteristic that it does not need to change the standardization and hardware of the existing wireless networks. The RIS is capable of smartly 'reconfiguring' the wireless propagation environment by compensating the power loss over long distances, as well as for forming virtual line-of-sight (LoS) links between the base stations (BSs) and the mobile users (MUs) via passively reflecting their received signal. The throughput enhancement becomes more considerable when the LoS link between BSs and MUs is blocked by high-rise buildings with high probability. Due to the intelligent deployment and design of the RIS, a software-defined wireless environment may be constructed, which in turn, provides potential received signal-to-interference-plus-noise ratio (SINR) enhancements. In contrast to the conventional relaying system, e.g., amplify-and-forward (AF) and decode-and-forward (DF), the RIS does not need a dedicated power source for operation, while it can be invoked with minimal hardware complexity~\cite{ntontin2019reconfigurable,bjornson2019intelligent}. Sparked by the aforementioned advantages, the application of RIS-enhanced communication networks is highly desired.
\vspace{-0.5cm}

\subsection{State-of-the-art}

Again, the RIS-enhanced wireless networks have attracted remarkable attention in recent years in diverse application scenarios. Several fundamental technical challenges are ready to be tackled, including the joint active beamforming for the BS and the passive beamforming for the RIS~\cite{ye2019joint,ding2019simple,wu2018intelligent,abeywickrama2019intelligent}, the deployment of the RIS~\cite{di2019smart}, hardware implementation~\cite{liaskos2018new,yang2016design}, as well as the channel modeling~\cite{lu2014overview,jensen2019optimal}.

\subsubsection{Joint active and passive design of the RIS-enhanced system}

To fully reap the benefits of the RIS in the wireless networks, joint active and passive design of the RIS-enhanced system have been considered in both multiple-input-single-output (MISO) scenarios and single-input-single-output (SISO) scenarios. The authors of~\cite{wu2018intelligent} jointly designed the active beamforming and passive beamforming to minimize the total transmit power under the user' SINR constraints in both single-user and multi-user scenarios. Simulation results in~\cite{wu2018intelligent} demonstrated that both the spectrum and energy efficiency were improved with the assistance of the RIS. The authors of~\cite{guo2019weighted} jointly designed the active beamforming and shifting the incident signal to discrete phase levels to maximize the weighted sum-rate of all MUs. The formulated non-convex problem in~\cite{guo2019weighted} was decoupled via Lagrangian dual transform, based on which, the sub-problem of active beamforming and passive beamforming were optimized in an iterative manner. The authors of~\cite{ye2019joint} also focused on optimizing the phase shift of the RIS, as well as the precoding matrix of the BS. Thus, the symbol error rate (SER) of the system was minimized. The sub-problem of reflecting and precoding was iteratively solved by minimizing the symbol error rate (MSER). The authors of~\cite{huang2019reconfigurable} aimed at maximizing the energy efficiency of the RIS-enhanced system by designing the phase shift of the RIS and the power allocation policy from the BS to the MUs. Additionally, a realistic energy consumption model of the RIS was presented. Finally, a realistic outdoor environment was invoked for analyzing the performance of the RIS-enhanced system. Simulation results in~\cite{huang2019reconfigurable} demonstrated that the network is capable of obtaining up to 300\% higher energy efficiency compared to conventional multi-antenna AF relaying.

\subsubsection{NOMA in the RIS-enhanced wireless networks}

In an effort to improve the spectrum efficiency and user connectivity of the RIS-enhanced wireless networks, power-domain non-orthogonal multiple access (NOMA) technology is adopted, whose key idea is to superimpose the signals of two MUs at different powers for exploiting the spectrum more efficiently by opportunistically exploring the users' different channel conditions~\cite{liu2017enhancing,liu2017non}. The authors of~\cite{li2019joint} considered a MISO-NOMA downlink communication network for minimizing the total transmit power by jointly designing the transmit precoding vectors and the reflecting coefficient vector. In~\cite{yang2019intelligent}, the authors jointly optimized the phase shift of the RIS, as well as the power allocation from the BS to the MUs. Thus, the minimum decoding SINR of all MUs was maximized for optimizing the throughput of the system with considering user fairness. The authors of~\cite{ding2019simple} proposed a simple design of RIS assisted NOMA transmission. It can be observed in~\cite{ding2019simple} that, the directions of users' channel vectors are capable of being aligned with the aid of the RIS, which emphasizes the importance of implementing NOMA technology. For the NOMA-assisted RIS system, the core challenge is that the decoding order is dynamically changed due to the phase shift of the RIS. The authors of~\cite{mu2019exploiting} proposed a NOMA-RIS-MISO framework to maximize the throughput of the system with considering the dynamic decoding order condition. Successive convex approximation (SCA) technique and sequential rank-one constraint relaxation (SROCR)-based algorithm were invoked for obtaining a locally optimal solution.

\subsubsection{Machine learning in the RIS-enhanced wireless networks}

As a benefit of the machine learning (ML) based framework, many challenges in the conventional wireless communication networks have been circumvented, leading to enhanced network performance, improved reliability and agile adaptivity~\cite{liu2019machine,jiang2016machine,qin2019deep}. At the time of writing, only deep learning (DL) is invoked in the RIS-enhanced wireless networks. The authors of~\cite{taha2019enabling} adopted the DL method for learning the reflection matrices of the RIS directly from the sampled channel knowledge without any knowledge of the RIS array geometry. The authors of~\cite{huang2019indoor} leveraged a deep neural network (DNN)-based approach in the indoor communication environment for estimating the mapping between a MU's position and the configuration of the RIS's unit cells. Thus, the received SNR was maximized. However, there is a paucity of research on invoking the ML methods for dynamically deploying and designing the RIS for achieving higher performance than the conventional approaches.
\vspace{-0.5cm}

\subsection{Motivations}

As mentioned above, research on the deployment of the RIS is fundamental but essential. However, there is a paucity of research on the problem of position determination of the RIS. Additionally, current research contributions mainly consider the performance optimization for both single-MU and multi-MU scenarios by optimizing the phase shift and/or pre-coding solutions of the RIS-empowered system~\cite{nadeem2019large,ye2019joint,liang2019large,jung2019performance,pan2019Intelligent}. Considering the deployment of the RIS based on the MU' particular data demand implicitly assumes that the long-term tele-traffic requirement of MUs is already known or it can be learned/predicted. With this proviso, the deployment and control method of the RIS may be designed periodically for maximizing the long-term benefits and hence reduce the additional control. Meanwhile, in an effort to maximize the MUs' satisfaction level, the RIS is supposed to learn by interacting with the environment and adapt the control/deployment policy based on the feedback of the MUs to overcome the uncertainty of the environment. To the best of our knowledge, this important problem is still unsolved. Additionally, the power dissipation of the RIS derives from the controlling of the varactor diodes (hold the functionality of controlling the phase shift). However, there is also a paucity of formulating the power consumption model based on the power dissipation of the varactor diodes. Finally, the next-generation wireless networks are expected to transmit several hundreds times more data than the current generation. However, the energy dissipation are expected to be the same level, which indicates that the next-generation wireless networks are supposed to be designed in cost-efficient and environmentally friendly manner~\cite{Jiayi2019Multiple}. Thus, energy efficiency is much more anticipated than throughput in some energy-limited application scenario, such as unmanned aerial vehicle (UAV)-enabled wireless networks. In this paper, we aim for maximizing energy efficiency instead of maximizing the throughput of the RIS-enhanced cellular networks. Therefore, the problem of joint deployment design and phase shift control of the RIS is formulated for maximizing the energy efficiency of the RIS, while satisfying the particular data requirement of each MU.

In terms of the methodology, reinforcement learning (RL) methods have also witnessed increasing applications in the fifth-generation (5G) wireless systems~\cite{liu2019machine}. The core idea of the RL-assisted techniques adopted in the RIS-enhanced wireless networks is that they allow the BS/RIS to improve their service quality by learning from the environment, from their historical experience and from the feedback of the MUs~\cite{liu2019trajectory}. More explicitly, RL models can be used for supporting the BS/RIS (agents) in their interaction with the environment (states) and by learning from their mistakes, whilst finding the optimal behavior (actions) of the BS/RIS. Furthermore, the RL model can incorporate farsighted system evolution (long-term benefits) instead of only focusing on current states. Thus, it is invoked for solving challenging problems in the RIS-enhanced wireless networks.
\vspace{-0.5cm}

\subsection{Contributions}

Against the aforementioned background, our new contributions are as follows:

\begin{itemize}
\item We propose a novel framework for the long-term control and deployment design of RIS-enhanced MISO-NOMA networks, in which the RIS is installed to enhance the wireless service. Based on the proposed framework, we formulate the energy efficiency maximization problem by jointly designing the phase shift, power allocation and position of the RIS. Additionally, a novel power dissipation model is formulated by considering the varactor diodes.
\item We adopt a novel long short-term memory (LSTM) based echo state network (ESN) algorithm, which is formed based on the architecture of an ESN model while leveraging LSTM units as hidden neurons, for obtaining users' tele-traffic demand with the aid of a real dataset.
\item We conceive a decaying double deep Q-network (${{\text{D}}^{\text{3}}}{\text{QN}}$) based algorithm for the joint position design and phase shift control problem of the RIS. In contrast to the conventional DQN algorithm, the ${{\text{D}}^{\text{3}}}{\text{QN}}$ algorithm is capable of overcoming the large overestimation of action values caused in the Q-learning model. Meanwhile, the performance of the ${{\text{D}}^{\text{3}}}{\text{QN}}$ algorithm can be further improved by leveraging the decaying $\epsilon $-greedy strategy.
\item We demonstrate that the proposed LSTM-based ESN algorithm is capable of striking a tradeoff between the prediction accuracy and computational complexity, while the proposed ${{\text{D}}^{\text{3}}}{\text{QN}}$ based algorithm outperforms the benchmarks in terms of energy efficiency.
\end{itemize}


The rest of the paper is organized as follows. In Section II, the problem formulation of energy efficiency for the RIS is presented. In Section III, the proposed DQN and ${{\text{D}}^{\text{3}}}{\text{QN}}$ based algorithms conceived for solving the problem formulated are demonstrated. Our numerical results are presented in Section V, which is followed by our conclusions in Section VI.

\vspace{-0.3cm}

\section{System Model and Problem Formulation}

\subsection{System Model}

As illustrated in Fig.1, the downlink MISO communications between a BS equipped with $M$ antenna elements and $K$ single-antenna MUs in a particular area are considered. A RIS with $N$ reflecting elements is installed on the facade of a building for enhancing the wireless services~\cite{di2019smart,qingqing2019towards}. The RIS is linked with a controller, which controls the reflecting elements for hosting the functionality of phase-shifting and amplitude absorption. NOMA technology is invoked for further enhancing the spectrum efficiency of the system~\cite{ding2019simple}. In the RIS-enhanced system, the MUs are partitioned into $L$ clusters, while the number of MUs in each cluster is assumed as two for the sake of simplicity. Since the received signal of each MU is composite, we need to distinguish the strong MU (having a larger channel gain) with the weak MU (having a smaller channel gain). It is also assumed that the channel state information (CSI) of all channels is perfectly known by the BS via ray tracing technology~\cite{yang2019intelligentglobecom,wu2018intelligent}.

\begin{figure} [t!]
\centering
\includegraphics[width=4.2in]{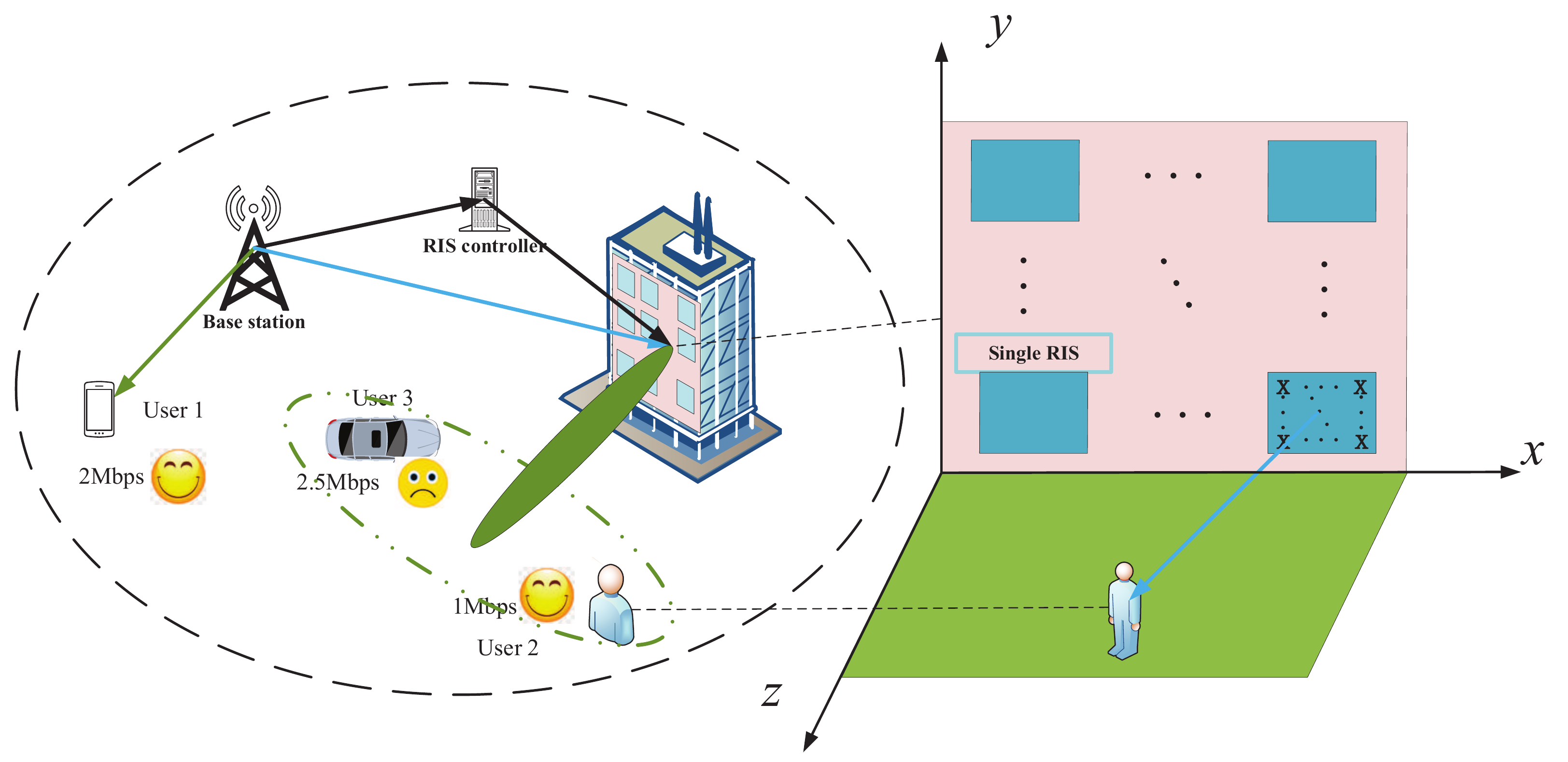}
\caption{Illustration of intelligent reflecting surface in wireless communications.}\label{scenario}
\end{figure}

Due to the employment of the RIS, the composite received signal is a concatenation of two components, namely signals derived from direct link between BS and MUs (BS-MU link), and signals derived from reflecting link (BS-RIS-MU link). With the aid of the RIS, a software-defined wireless environment is constructed, which in turn, obtains potential enhancements of received SINR. The three individual channels, namely BS-RIS link, RIS-MU link, and BS-MU link are denoted by ${\boldsymbol{H}}_{B,S} \in {\mathbb{C}^{N \times M}}$, $\boldsymbol{h}_{S,k}^{H} \in {\mathbb{C}^{1 \times N}}$, and $\boldsymbol{h}_{B,k}^{H} \in {\mathbb{C}^{1 \times M}}$, respectively, with $k \in \mathcal{K},{\kern 1pt} {\kern 1pt} {\kern 1pt} \left| \mathcal{K} \right| = K$ and $n \in \mathcal{N},{\kern 1pt} {\kern 1pt} {\kern 1pt} \left| \mathcal{N} \right| = N$, while ${{\boldsymbol{h}}^H}$ denotes the conjugate transpose of matrix $\boldsymbol{h}$. In terms of the RIS, denote ${{\boldsymbol{\theta}}_S} = [{\theta _{S,1}}, \cdots ,{\theta _{S,n}}, \cdots ,{\theta _{S,N}}]$, then, ${{\mathbf{\Theta }}_S} = {\text{diag}}\left( {{\beta _{S,1}}{e^{j{\theta _{S,1}}}}, \cdots ,{\beta _{S,n}}{e^{j{\theta _{S,n}}}}, \cdots ,{\beta _{S,N}}{e^{j{\theta _{S,N}}}}} \right)$ represents the diagonal phase-shifting matrix of the RIS, where ${\theta _{S,n}} \in \left[ {0,2\pi } \right]$ represents the phase shift, while ${\beta _{S,n}} \in \left[ {0,1} \right]$ denotes the amplitude reflection coefficient~\cite{guo2019weighted}\footnote[1]{We assume ${\beta _{S,n}}=1$ in this paper, since it is energy-costly to implement independent control of the reflecting amplitude and phase shift simultaneously. The relationship of the phase shift and amplitude can be found in~\cite{abeywickrama2019intelligent}.}.

\subsubsection{Zero-forcing precoding method}

Denote ${x_l}(t) = \sqrt {{\alpha _{l,a}}(t)} {s_{l,a}}(t) + \sqrt {{\alpha_{l,b}}(t)} {s_{l,b}}(t)$ as the transmit signal of the $l$-th cluster, while ${s_{l,a}}$ and ${s_{l,b}}$ are the signals for MU $a$ and MU $b$ in the $l$-th cluster, respectively, while ${\alpha_{l,a}}$ and ${\alpha_{l,b}}$ are the power allocation factors for MU $a$ and MU $b$, respectively, and $\alpha_{l,a}+\alpha_{l,b}=1$. It is worth noting that, all equations in this paper are time-varying, we omit the $(t)$ in equations of this paper for expression simplify. Hence, the received signal of a MU is given by
\begin{align}\label{transmitsignal}
{{y_{l,i}} = \left( {{\boldsymbol{h}}_{B,l,i}^H + {\boldsymbol{h}}_{S,l,i}^H{{\boldsymbol{\Theta }}_S}{{\boldsymbol{H}}_{B,S}}} \right)\sum\limits_{l = 1}^L {{{\boldsymbol{w}}_l}{x_l}}  + {n_{l,i}} },
\end{align}
where ${n_{l,i}} \sim \mathcal{C}\mathcal{N}\left( {0,\sigma _{l,i}^2} \right)$ represents the additive white Gaussian noise (AWGN) at the MU's receiver. Finally, ${{{\boldsymbol{w}}_k}}$ denotes the corresponding beamforming vector of the $l$-th cluster.

In a particular cluster, each MU tries to employ successive interference cancelation (SIC) in a successive order to remove the intra-beam interference~\cite{liu2016cooperative,ding2017survey}. MU $a$ ($b$) is named as strong (weak) user. The strong MU can remove the interference from the weak MU by SIC, while the weak user decodes the received signal directly without SIC~\cite{liu2018multiple,liu2017non}. Based on equation \eqref{transmitsignal}, the received signal of the strong user in the $l$-th cluster is given by
\begin{align}\label{SINR}
{\begin{gathered}
  {y_{l,a}} =  \left( {{\boldsymbol{h}}_{B,l,a}^H + {\boldsymbol{h}}_{S,l,a}^H{{\boldsymbol{\Theta }}_S}{{\boldsymbol{H}}_{B,S}}} \right){{\boldsymbol{w}}_l}\left( {\sqrt {{\alpha _{l,a}}} {s_{l,a}} + \sqrt {{\alpha_{l,b}}} {s_{l,b}}} \right) \hfill \\
  {\kern 1pt} {\kern 1pt} {\kern 1pt} {\kern 1pt} {\kern 1pt} {\kern 1pt} {\kern 1pt} {\kern 1pt} {\kern 1pt} {\kern 1pt} {\kern 1pt} {\kern 1pt} {\kern 1pt} {\kern 1pt} {\kern 1pt} {\kern 1pt} {\kern 1pt} {\kern 1pt} {\kern 1pt} {\kern 1pt} {\kern 1pt} {\kern 1pt} {\kern 1pt} {\kern 1pt} {\kern 1pt} {\kern 1pt} {\kern 1pt} {\kern 1pt} {\kern 1pt}  + \left( {{\boldsymbol{h}}_{B,l,a}^H(t) + {\boldsymbol{h}}_{S,l,a}^H{{\boldsymbol{\Theta }}_S}{{\boldsymbol{H}}_{B,S}}} \right)\sum\limits_{j = 1,j \ne l}^L {{{\boldsymbol{w}}_j}{x_j}}  + {n_{l,a}}, \hfill \\
\end{gathered} }
\end{align}
where $\left( {{\boldsymbol{h}}_{B,l,a}^H + {\boldsymbol{h}}_{S,l,a}^H{{\boldsymbol{\Theta }}_S}{{\boldsymbol{H}}_{B,S}}} \right)\sum\limits_{j = 1,j \ne l}^L {{{\boldsymbol{w}}_j}{x_j}} $ is the inter-cluster interference from other clusters, while $\left( {{\boldsymbol{h}}_{B,l,a}^H + {\boldsymbol{h}}_{S,l,a}^H{{\boldsymbol{\Theta }}_S}{{\boldsymbol{H}}_{B,S}}} \right){{\boldsymbol{w}}_l}\sqrt {{a_{l,b}}} {s_{l,b}}$ is the intra-cluster interference from the weak user in the same cluster.

To eliminate the inter-cluster interference, zero-forcing (ZF)-based linear pre-coding method is employed~\cite{huang2019reconfigurable,basar2019large,huang2018energy}. Although the dirty paper coding (DPC) is proved to achieve the maximum capacity in multi-user MIMO-NOMA system~\cite{liu2018multiple,liu2017non}, it is non-trivial to be implemented in practice for the reason that it adopts brute-force searching. Denote ${\boldsymbol{h}}_j^H,{\kern 1pt} {\kern 1pt} {\kern 1pt} j \in \mathcal{L}$ as the combined channel of the $l$-th cluster, the corresponding ZF pre-coding constraints are given by
\begin{align}\label{ZF1}
{\left\{ {\begin{array}{*{20}{c}}
  {\left[ {{\boldsymbol{h}}_{B,j}^{H} + {\boldsymbol{h}}_{S,j}^{H}{{\boldsymbol{\Theta }}_S}{\boldsymbol{H}}_{B,S}} \right]{{\boldsymbol{w}}_l} = 0,}&{{\kern 1pt} \forall j \ne l,{\kern 1pt} {\kern 1pt} {\kern 1pt} j \in \mathcal{L},} \\
  {\left[ {{\boldsymbol{h}}_{B,l}^{H} + {\boldsymbol{h}}_{S,l}^{H}{{\boldsymbol{\Theta }}_S}{\boldsymbol{H}}_{B,S}} \right]{{\boldsymbol{w}}_l} = 1,}&{j = l.}
\end{array}} \right.}
\end{align}


Denote ${{\boldsymbol{H}}^H} = {\boldsymbol{H}}_{BM}^H + {\boldsymbol{H}}_{SM}^H{{\boldsymbol{\Theta }}_S}{\boldsymbol{H}}_{BS}$, where we have ${\boldsymbol{H}}_{BM}^H = {\left[ {{\boldsymbol{h}}_{B,1,a}, \cdots ,{\boldsymbol{h}}_{B,L,a}} \right]^H}$ and ${\boldsymbol{H}}_{SM}^H = {\left[ {{\boldsymbol{h}}_{S,1,a}, \cdots ,{\boldsymbol{h}}_{S,L,a} } \right]^H}$. Thus, the optimal transmit pre-coding metric ${\boldsymbol{W}} = \left[ {{{\boldsymbol{w}}_1}, \cdots ,{{\boldsymbol{w}}_L}} \right]$ is given by the pseudo-inverse of the combined channel ${{\boldsymbol{H}}^H}$ as follows
\begin{align}\label{ZF2}
{{\boldsymbol{W}} = {\boldsymbol{H}}{\left( {{{\boldsymbol{H}}^H}{\boldsymbol{H}}} \right)^{ - 1}}}.
\end{align}

\begin{remark}\label{remark:ZFdrawback}
The NOMA-ZF beamforming vector is generated by the channel of the strong MUs in each cluster. Thus, the weak MUs are affected by the interferences from both other clusters and the strong MU in the same cluster.
\end{remark}

Following the equation \eqref{ZF2}, the strong MUs is capable of removing the inter-cluster interference by NOMA-ZF beamforming and removing the intra-user interference by employing SIC. Thus, the received signal of the strong MU in the $l$-th cluster is given by
\begin{align}\label{yl1}
{{y_{l,a}}{\kern 1pt}  = \left( {{\boldsymbol{h}}_{B,l,a}^H + {\boldsymbol{h}}_{S,l,b}^H{{\boldsymbol{\Theta }}_S}{{\boldsymbol{H}}_{B,S}}} \right){{\boldsymbol{w}}_l}\sqrt {{\alpha_{l,a}}} {s_{l,a}} + {n_{l,a}}}.
\end{align}

Thus, the received SINR of the strong MU in the $l$-th cluster is given by
\begin{align}\label{SINR1}
{{\gamma _{l,a}} = \frac{{{{\left| {\left( {{\boldsymbol{h}}_{B,l,a}^H + {\boldsymbol{h}}_{S,l,a}^H{{\boldsymbol{\Theta }}_S}{{\boldsymbol{H}}_{B,S}}} \right){{\boldsymbol{w}}_l}\sqrt {{\alpha_{l,a}}} {s_{l,a}}} \right|}^2}}}{{\sigma _l^2}} = \frac{{{\alpha_{l,a}}{P_{l}}}}{{\sigma _l^2}} }.
\end{align}

The received signal of the weak MU in the $l$-th cluster can be expressed as
\begin{align}\label{yl2}
{\begin{gathered}
  {y_{l,b}}{\kern 1pt}  = \left( {{\boldsymbol{h}}_{B,l,b}^H + {\boldsymbol{h}}_{S,l,b}^H{{\boldsymbol{\Theta }}_S}{{\boldsymbol{H}}_{B,S}}} \right){{\boldsymbol{w}}_l}\left( {\sqrt {{\alpha_{l,a}}} {s_{l,a}} + \sqrt {{\alpha_{l,b}}} {s_{l,b}}} \right) \hfill \\
  {\kern 1pt} {\kern 1pt} {\kern 1pt} {\kern 1pt} {\kern 1pt} {\kern 1pt} {\kern 1pt} {\kern 1pt} {\kern 1pt} {\kern 1pt} {\kern 1pt} {\kern 1pt} {\kern 1pt} {\kern 1pt} {\kern 1pt} {\kern 1pt} {\kern 1pt} {\kern 1pt} {\kern 1pt} {\kern 1pt} {\kern 1pt} {\kern 1pt} {\kern 1pt} {\kern 1pt} {\kern 1pt} {\kern 1pt} {\kern 1pt} {\kern 1pt} {\kern 1pt}  + \left( {{\boldsymbol{h}}_{B,l,b}^H + {\boldsymbol{h}}_{S,l,b}^H{{\boldsymbol{\Theta }}_S}{{\boldsymbol{H}}_{B,S}}} \right)\sum\limits_{j = 1,j \ne l}^L {{{\boldsymbol{w}}_j}{x_j}}  + {n_{l,b}}. \hfill \\
\end{gathered} }
\end{align}

It can be observed that the inter-cluster interference for the weak MU is not capable of being eliminate by leveraging ZF beamforming, while the weak MU does not perform SIC. Therefore, the received SINR of the weak MU in the $l$-th cluster is given by
\begin{align}\label{SINR2}
{{\gamma _{l,b}} = \frac{{{{\left| {{{\boldsymbol{h}}_{l,b}}{{\boldsymbol{w}}_l}} \right|}^2}{\alpha_{l,b}}{P_{l}}}}{{{{\left| {{{\boldsymbol{h}}_{l,b}}{{\boldsymbol{w}}_l}} \right|}^2}{\alpha_{l,b}}{P_{l}} + {{\left| {{{\boldsymbol{h}}_{l,b}}\sum\limits_{j = 1,j \ne l}^L {{{\boldsymbol{w}}_l}{x_j}} } \right|}^2} + \sigma _l^2}} },
\end{align}
where we have ${{\boldsymbol{h}}_{l,b}} = {\boldsymbol{h}}_{B,l,b}^H+ {\boldsymbol{h}}_{S,l,b}^H{{\boldsymbol{\Theta }}_S}{{\boldsymbol{H}}_{B,S}}$.

\subsubsection{projection hybrid NOMA precoding method}

Since ZF precoding method is a single beam per group (SBPG) approach~\cite{chen2016beamforming,choi2015minimum}, if the data demand of the strong MU is comparable to that of the weak MU, the power allocation factor $a_{l,2}$ has to be sufficiently close to 0 in terms of fairness. Thus, a lower spectral efficiency is suffered. a project hybrid NOMA (PH-NOMA) precoding method~\cite{chen2016beamforming}, which combines the conventional ZF method and hybrid NOMA method, is invoked as the precoding approach in this paper for removing inter-cluster interference.

Denote $\widehat {\boldsymbol{H}}_{l,a,b}^H = \left( {{{\boldsymbol{H}}_1}, \cdots ,{{\boldsymbol{H}}_{b - 1}},{{\boldsymbol{H}}_{b + 1}}, \cdots ,{{\boldsymbol{H}}_{a - 1}},{{\boldsymbol{H}}_{a + 1}}, \cdots ,{{\boldsymbol{H}}_K}} \right)$, which indicates that $\widehat {\boldsymbol{H}}_{l,a,b}^H$ is the sub-matrix obtained by striking ${{{\boldsymbol{H}}_{l,a}}}$ and ${{{\boldsymbol{H}}_{l,b}}}$ out of ${{\boldsymbol{H}}^H}$,. Note that $\widehat {\boldsymbol{H}}_{l,a,b}^H \in {\boldsymbol{C}^{M \times (K - 2)}}$. Then the orthogonal projection $\mathcal{P}_{l,a,b}^ \bot $ of $\widehat {\boldsymbol{H}}_{l,a,b}^H$ can be expressed as
\begin{align}\label{projection}
{\mathcal{P}_{l,a,b}^ \bot  = {{\boldsymbol{I}}_M} - \widehat {\boldsymbol{H}}_{l,a,b}^H{\left( {{{\widehat {\boldsymbol{H}}}_{l,a,b}}\widehat {\boldsymbol{H}}_{l,a,b}^H} \right)^{ - 1}}{\widehat {\boldsymbol{H}}_{l,a,b}}},
\end{align}
with $\mathcal{P}_{l,a,b}^ \bot {{\boldsymbol{H}}_j} = 0,{\kern 1pt} {\kern 1pt} {\kern 1pt} {\kern 1pt} {\kern 1pt} {\kern 1pt} {\kern 1pt} {\kern 1pt} j \ne a,b$.

The superposition for transmitting $s_{l,a}$ and $s_{l,b}$ can be formulated as
\begin{align}\label{PHxl}
{{{\boldsymbol{x}}_l} = \mathcal{P}_{l,a,b}^ \bot \left( {{{\boldsymbol{w}}_{l,a}}{s_{l,a}} + {{\boldsymbol{w}}_{l,b}}{s_{l,b}}} \right),{\boldsymbol{H}}_j^H{\boldsymbol{x_l}} = 0,{\kern 1pt} {\kern 1pt} {\kern 1pt} {\kern 1pt} {\kern 1pt} {\kern 1pt} {\kern 1pt} j \ne a,b}.
\end{align}

Note that ${\boldsymbol{H}}_j^H{\boldsymbol{x_l}} = 0,{\kern 1pt} {\kern 1pt} {\kern 1pt} {\kern 1pt} {\kern 1pt} {\kern 1pt} {\kern 1pt} j \ne a,b$ since $\mathcal{P}_{l,a,b}^ \bot $ is the orthogonal projection of $\widehat {\boldsymbol{H}}_{l,a,b}^H$. Therefore, the received signal of user $a$ in the $l$-th cluster can be expressed as
\begin{align}\label{PHya}
{{y_{l,a}} = {\boldsymbol{H}}_{l,a}^H\mathcal{P}_{l,a,b}^ \bot \left( {{{\boldsymbol{w}}_{l,a}}{s_{l,a}} + {{\boldsymbol{w}}_{l,b}}{s_{l,b}}} \right) + {n_{l,a}} = {\left( {\mathcal{P}_{l,a,b}^ \bot {{\boldsymbol{H}}_{l,a}}} \right)^H}\left( {{{\boldsymbol{w}}_{l,a}}{s_{l,a}} + {{\boldsymbol{w}}_{l,b}}{s_{l,b}}} \right) + {n_{l,a}}}.
\end{align}

Similarly, the the received signal of user $b$ in the $l$-th cluster is given by
\begin{align}\label{PHyb}
{{y_{l,b}} = {\boldsymbol{H}}_{l,b}^H\mathcal{P}_{l,a,b}^ \bot \left( {{{\boldsymbol{w}}_{l,a}}{s_{l,a}} + {{\boldsymbol{w}}_{l,b}}{s_{l,b}}} \right) + {n_{l,b}} = {\left( {\mathcal{P}_{l,a,b}^ \bot {{\boldsymbol{H}}_{l,b}}} \right)^H}\left( {{{\boldsymbol{w}}_{l,a}}{s_{l,a}} + {{\boldsymbol{w}}_{l,b}}{s_{l,b}}} \right) + {n_{l,b}}}.
\end{align}

Thus, the received SINR for user $a$ and user $b$ in the $l$-th cluster can be expressed as
\begin{align}\label{SINEab}
{{\begin{array}{*{20}{c}}
  {{\gamma _{l,a}} = \frac{{{{\left| {{\boldsymbol{H}}_{l,a}^H\mathcal{P}_{l,a,b}^ \bot {{\boldsymbol{w}}_{l,a}}{s_{l,a}}} \right|}^{\text{2}}}}}{{\sigma _{l,a}^2}},} \\
  {{\gamma _{l,b}} = \frac{{{{\left| {{\boldsymbol{H}}_{l,b}^H\mathcal{P}_{l,a,b}^ \bot {{\boldsymbol{w}}_{l,b}}{s_{l,b}}} \right|}^{\text{2}}}}}{{{{\left| {{\boldsymbol{H}}_{l,b}^H\mathcal{P}_{l,a,b}^ \bot {{\boldsymbol{w}}_{l,a}}{s_{l,a}}} \right|}^{\text{2}}} + \sigma _{l,b}^2}},}
\end{array}}}
\end{align}
with the constraint of ${\gamma}{_{l,b \to l,a}} \ge {\gamma}_{l,b \to l,b}$~\cite{cui2017optimal}, where $\gamma_{l,b \to l,a}$ represents the SINR of user $a$ to decode user $b$. The responding decoding rate is ${R_{l,b \to l,a}} = {\log _2}\left( {1 + \gamma_{l,b \to l,a}} \right)$. Under the assumption of a given decoding order, to guarantee SIC performed successfully, the condition ${R_{l,b \to l,a}} \geqslant {R_{l,b \to l,b}}$ for ${\pi _l}(a) \ge {\pi _l}(b)$, while ${\pi _l}(a)=1,{\pi _l}(b)=2$ is the decoding order.

In the scenario that three users $a,b,c$ are partitioned into the $l$-th cluster, the SIC decoding order constraint can be given by
\begin{align}\label{decodingorder}
{{R_{l,b \to l,a}} \geqslant {R_{l,b \to l,b}},{\kern 1pt} {\kern 1pt} {\kern 1pt} {\kern 1pt} {\kern 1pt} {\kern 1pt} {\kern 1pt} {\kern 1pt} {\kern 1pt} {\kern 1pt} {R_{l,c \to l,a}} \geqslant {R_{l,c \to l,c}},{\kern 1pt} {\kern 1pt} {\kern 1pt} {\kern 1pt} {\kern 1pt} {\kern 1pt} {\kern 1pt} {\kern 1pt} {\kern 1pt} {\kern 1pt} {R_{l,c \to l,b}} \geqslant {R_{l,c \to l,c}}}.
\end{align}

By applying Propositions 1 and Propositions 2 in~\cite{chen2016beamforming}, we can obtain
\begin{align}\label{PHwab}
{{{\boldsymbol{w}}_{l,a}} = {\nu _{l,a}}\left( {\left( {1 + {R_{{{\min }^{l,b}}}}} \right){{\mathbf{e}}_{l,a}} - {R_{{{\min }^b}}}{\mathbf{e}}_{l,a}^H{{\mathbf{e}}_{l,a}}{{\mathbf{e}}_{l,b}}} \right),{\kern 1pt} {\kern 1pt} {\kern 1pt} {\kern 1pt} {\kern 1pt} {{\boldsymbol{w}}_{l,b}} = {\nu _{l,b}}{{\mathbf{e}}_{l,b}},}
\end{align}
where
\begin{align}\label{PHwab2}
{\left\{ {\begin{array}{*{20}{c}}
  {{{\boldsymbol{e}}_{l,a}} = \frac{{\mathcal{P}_{l,a,b}^ \bot {{\boldsymbol{H}}_{l,a}}}}{{\left\| {\mathcal{P}_{l,a,b}^ \bot {{\boldsymbol{H}}_{l,a}}} \right\|}},{{\boldsymbol{e}}_{l,b}} = \frac{{\mathcal{P}_{l,a,b}^ \bot {{\boldsymbol{H}}_{l,b}}}}{{\left\| {\mathcal{P}_{l,a,b}^ \bot {{\boldsymbol{H}}_{l,b}}} \right\|}},} \\
  {\nu _{l,a}^2 = \frac{{{R_{\min }^{l,a}}}}{{{{\left\| {\mathcal{P}_{l,a,b}^ \bot {{\boldsymbol{H}}_{l,a}}} \right\|}^2}}}\frac{1}{{{{\left( {1 + {R_{\min }^{l,b}}{{\sin }^2}\varphi } \right)}^2}}},} \\
  {\nu _{l,b}^2 = \frac{{{R_{\min }^{l,b}}}}{{{{\left\| {\mathcal{P}_{l,a,b}^ \bot {{\boldsymbol{H}}_{l,b}}} \right\|}^2}}} + \frac{{{R_{\min }^{l,a}}}}{{{{\left\| {{{\boldsymbol{H}}_{l,a}}} \right\|}^2}}}\frac{{{R_{\min }^{l,b}}{{\cos }^2}\varphi }}{{{{\left( {1 + {R_{\min }^{l,b}}{{\sin }^2}\varphi } \right)}^2}}}.}
\end{array}} \right.}
\end{align}

Note that $u = {\cos ^2}\varphi  = \frac{{{\boldsymbol{H}}_{l,b}^H{{\boldsymbol{H}}_a}{\boldsymbol{H}}_{l,a}^H{{\boldsymbol{H}}_{l,b}}}}{{{{\left\| {{{\boldsymbol{H}}_{l,a}}} \right\|}^2}{{\left\| {{{\boldsymbol{H}}_{l,b}}} \right\|}^2}}}$ represents the channel correlation between ${{\boldsymbol{H}}_{l,a}}$ and ${{\boldsymbol{H}}_{l,b}}$. Thus, the instantaneous transmit rate of user $ i \in \{ a,b\} $ in the $l$-th cluster is given by
\begin{align}\label{transmit rate}
{{R_{l,i}} = {B_l}{\log _2}\left( {1 + {\gamma _{l,i}}} \right)}.
\end{align}

\begin{remark}\label{remark:ZFNOMA}
In contrast to the ZF method, the PH-NOMA precoding method considers the beamforming for both two users in the same cluster. Thus, the sum achievable transmit rate of the two users can be increased.
\end{remark}

\vspace{-0.7cm}
\subsection{Channel Model}

Consider a dense urban area, where MUs are surrounded by a number of buildings as illustrated in Fig.1. Denote the position of the $k$-th MU as ${c_k^U} = {[{x_k},{y_k}]^T} \in {\mathbb{R}^{2 \times 1}},k \in \mathcal{K}$, while the coordinate of the BS is denoted as ${c_B} = {\left[ {{x_B},{y_B},{h_B}} \right]^T}$ with $h_B$ representing the height of the BS. Furthermore, the position of the RIS is denoted as $C = {[x_l,y_l,z_l]^T}$.


The distance-dependent channel path loss is modeled as $\eta \left( d \right) = {C_0}{\left( {\frac{d}{{{d_0}}}} \right)^{ - \alpha }}$~\cite{wu2018intelligent}, where $C_0$ represents the path loss in the condition of $d_0=1$ meters (m), $d$ denotes the link distance, and $\alpha$ is the path loss exponent. In terms of the small scale fading, we assume a Rayleigh fading channel model for the BS-MU and RIS-MU channels, while a Rician fading model for the BS-RIS channel.

\vspace{-0.5cm}

\subsection{Data Demand Prediction}
Before the RIS is deployed, the BS is supposed to estimate the data demand in its cellular network and identify possible congestion events. These requirements, in turn, motivate a predictive approach to RIS' deployment. To this end, the BS needs to leverage machine learning (ML) techniques to predict the downlink cellular traffic.

Time-division duplexing (TDD) protocol is leveraged for uploading the satisfaction levels from MUs to BS. We consider five hypotheses in relation to the five alternatives (on the ordinal scale) for each satisfaction state, the measurement is the mean opinion score (MOS) received by MUs. These are: excellent (4.5), good ($3.5 \sim 4.5$), fair ($2 \sim 3.5$), poor ($1 \sim 2$) and bad (1).

Resorting to the well known throughput to MOS mapping in~\cite{Cui2018TWC}, the above satisfaction levels can be determined as
\begin{align}\label{MOS }
{{Q_k} = {\lambda _k}{\log _{10}}\left( {{\tau _k}{R_k}} \right)}
\end{align}
where $\lambda$, $\tau $ are parameters depend on specific maximal and minimal throughput demand of MUs.


\vspace{-0.5cm}

\subsection{Power Dissipation Model}

Since the RIS is equipped with passive reflecting elements, where power-hungry active components are avoided, it is anticipated to consume far less power than the traditional AF relay system. The total energy dissipated for operating the RIS-enhanced system is composed of four parts, namely, the transmit power of the BS, the hardware energy dissipated in the BS, the hardware energy dissipated in the MUs'devices, as well as the energy dissipated for controlling the RIS~\cite{huang2019reconfigurable,Ribeiro2018energy}. Thus the power consumption of the services from BS to $2L$ MUs with the aid of the RIS can be expressed as $P = \sum\limits_{l = 1}^L {{P_l} + K \cdot {P_{{\text{MU}}}}}  + {P_{{\text{BS}}}} + {P_{{\text{RIS}}}}$, where ${{P_l}}$ represents the total transmit power at the BS for the $l$-th cluster, ${{P_{{\text{MU}}}}}$ denotes the hardware energy dissipated in the $k$-th MU device, while ${P_{{\text{BS}}}}$ and $P_{{\text{RIS}}}$ represent the total hardware energy dissipated at the BS and the power dissipation at the RIS, respectively.

Denote ${\boldsymbol{P}} = {\text{diag}}\left( {{P_1}, \cdots ,{P_L}} \right)$, and ${\boldsymbol{W}} = \left[ {{{\boldsymbol{w}}_1}, \cdots ,{{\boldsymbol{w}}_K}} \right] \in {\mathbb{C}^{M \times K}}$. Thus, the total transmission power dissipation from the BS to the MUs is given by $\sum\limits_{l = 1}^L {{P_l} = }\sum\limits_{l = 1}^L {\left( {{{\left\| {{{\boldsymbol{w}}_{l,a}}} \right\|}^2} + {{\left\| {{{\boldsymbol{w}}_{l,b}}} \right\|}^2}} \right)}$ with the transmit power consumed for MU $a$ and $b$ in the $l$-th cluster can be expressed as

\begin{align}\label{PBSa}
{\begin{gathered}
  {P_{l,a}} = {\left\| {{{\boldsymbol{w}}_{l,a}}} \right\|^2} = {\left\| {{\nu _{l,a}}\left( {\left( {1 + {R _{\text{min}}^{l,b}}} \right){{\boldsymbol{e}}_{l,a}} - {\gamma _{l,b}}{\boldsymbol{e}}_{l,b}^H{{\boldsymbol{e}}_{l,a}}{{\boldsymbol{e}}_{l,b}}} \right)} \right\|^2} \hfill \\
  {\kern 1pt} {\kern 1pt} {\kern 1pt} {\kern 1pt} {\kern 1pt} {\kern 1pt} {\kern 1pt}{\kern 1pt} {\kern 1pt} {\kern 1pt} {\kern 1pt} {\kern 1pt} {\kern 1pt} {\kern 1pt} {\kern 1pt} {\kern 1pt} {\kern 1pt}  = \frac{{{R _{\text{min}}^{l,a}}}}{{{{\left\| {\mathcal{P}_{l,a,b}^ \bot {{\boldsymbol{H}}_{l,a}}} \right\|}^2}}}\frac{{\left( {1 + {R _{\text{min}}^{l,b}}{{\sin }^2}\varphi } \right)\left( {1 + {R _{\text{min}}^{l,b}}} \right) - {R _{\text{min}}^{l,b}}{{\cos }^2}\varphi }}{{{{\left( {1 + {R _{\text{min}}^{l,b}}{{\sin }^2}\varphi } \right)}^2}}}, \hfill \\
   {P_{l,b}} = {\left\| {{{\boldsymbol{w}}_{l,b}}} \right\|^2} = {\left\| {{\nu _{l,b}}} \right\|^2} = \frac{{{R _{\text{min}}^{l,b}}}}{{{{\left\| {\mathcal{P}_{l,a,b}^ \bot {{\boldsymbol{H}}_{l,b}}} \right\|}^2}}} + \frac{{{R _{\text{min}}^{l,b}}}}{{{{\left\| {\mathcal{P}_{l,a,b}^ \bot {{\boldsymbol{H}}_{l,a}}} \right\|}^2}}}\frac{{{R _{\text{min}}^{l,b}}{{\cos }^2}\varphi }}{{{{\left( {1 + {R _{\text{min}}^{l,b}}{{\sin }^2}\varphi } \right)}^2}}}. \hfill \\
\end{gathered}  }
\end{align}


Noted that the power consumption of the RIS derives from the power dissipation of the varactor diodes (The diode whose internal capacitance varies with the variation of the reverse voltage such type of diode is known as the Varactor diode). Therefore, the power dissipated at the RIS can be expressed as ${P_{{\text{RIS}}}}(t)= NP_n$, where $P_n$ represents the power dissipation of each varactor diode. Thus the total amount of power consumption of the RIS-enhanced system is given by ${P} = \sum\limits_{l = 1}^L {\left( {{{\left\| {{{\boldsymbol{w}}_{l,a}}} \right\|}^2} + {{\left\| {{{\boldsymbol{w}}_{l,b}}} \right\|}^2}} \right)}  + K{P_{{\text{MU}}}} + {P_{{\text{BS}}}} + N{P_n}$.



\subsection{Problem Formulation}
In this paper, we will design a protocol for controlling the RIS to assist/supplement the cellular networks. We are interested in maximizing the energy efficiency of the considered RIS-enhanced system. This performance is defined as the ratio between the system achievable sum MOS and the sum energy dissipation in Joule, which is given by
\begin{align}\label{EE}
{ \begin{gathered}
  \overline {{\eta}}_{EE}  = \frac{1}{T}\sum\limits_{t = 0}^T {\left( {\frac{{\sum\limits_{l = 1}^L {\left[ {{Q_{l,a}}\left( t \right) + {Q_{l,b}}\left( t \right)} \right]} }}{{\sum\limits_{l = 1}^L {\left( {{{\left\| {{{\boldsymbol{w}}_{l,a}}\left( t \right)} \right\|}^2} + {{\left\| {{{\boldsymbol{w}}_{l,b}}\left( t \right)} \right\|}^2}} \right)}  + K{P_{{\text{MU}}}}\left( t \right) + {P_{{\text{BS}}}}\left( t \right) + N{P_n}\left( t \right)}}} \right)}  \hfill \\
\end{gathered}  }.
\end{align}

\begin{remark}\label{remark:trend}
It can be observed from MUs' achievable SINR that, adding the transmit power or the number of reflecting elements leads to a higher received SINR. However, since the maximal MOS of the MUs is fixed, once the transmit power and the number of reflecting elements are high enough, increasing these two parameters leads to the reduction of energy efficiency.
\end{remark}

Denote ${{\boldsymbol{\theta}}} = [{\theta _{1}}, \cdots ,{\theta _{n}}, \cdots ,{\theta _{N}}]$, ${\boldsymbol{P}} = \left[ {{P_1}, \cdots ,{P_L}} \right]$, and $C = {[x_l,y_l,z_l]^T}$. We are interested in maximizing the long-term energy efficiency by optimizing the phase shift and three-dimensional (3D) position of the RIS, as well as the power allocation from the BS to MUs. Additionally, the dynamic decoding order has to be determined for guaranteeing SIC performance successfully. Thus, the optimization problem is formulated as

\begin{center}
\begin{subequations}\label{optimizationproblem2}
\begin{align}
{\mathop {\max }\limits_{\boldsymbol{\theta}, \boldsymbol{P}, \boldsymbol{\pi }, \boldsymbol{C}} {\kern 1pt} {\kern 1pt} {\kern 1pt}  \frac{1}{T}\sum\limits_{t = 0}^T {\left( {\frac{{\sum\limits_{l = 1}^L {\left[ {{Q_{l,a}}\left( t \right) + {Q_{l,b}}\left( t \right)} \right]} }}{{\sum\limits_{l = 1}^L {\left( {{{\left\| {{{\mathbf{w}}_{l,a}}\left( t \right)} \right\|}^2} + {{\left\| {{{\mathbf{w}}_{l,b}}\left( t \right)} \right\|}^2}} \right)}  + K{P_{{\text{MU}}}}\left( t \right) + {P_{{\text{BS}}}}\left( t \right) + N{P_n}\left( t \right)}}} \right)} } \\
{\text{s}}{\text{.t}}{\text{.}}{\kern 1pt} {\kern 1pt} {\kern 1pt} {\kern 1pt} {\kern 1pt} {\kern 1pt} {\kern 1pt} {\kern 1pt} {\kern 1pt} {\kern 1pt} {\kern 1pt} {\kern 1pt} {\kern 1pt} {\kern 1pt} {\kern 1pt} {\kern 1pt} {\kern 1pt} {\kern 1pt} {R_{l,i}}(t) \ge R_{\min }^{l,i}(t) ,\forall k, \forall l, \forall i \in \{ a,b\}, \\
0 \le {\theta _{l,n}}(t) \le \pi ,\forall l,\forall n, \\
c_l^I \in c_m^O,\forall l,\forall m, \\
{R_{l,b \to l,a}} (t)  \ge {R_{l,b \to l,b}} (t), {\pi _l}(a) \ge {\pi _l}(b) \forall l, \\
\sum\limits_{l = 1}^L {\left( {{{\left\| {{{\boldsymbol{w}}_{l,a}}} \right\|}^2} + {{\left\| {{{\boldsymbol{w}}_{l,b}}} \right\|}^2}} \right)} \le {P_{\max }} ,\forall k,
\end{align}
\end{subequations}
\end{center}
where $R_{\min }^{l,i}(t)$ is the minimal average achievable rate of user $i$ in cluster $l$ at timeslot $t$. Therefore, (21b) represents the transmit rate constraint in consideration of fairness. (21c) denotes the phase shift constraint of the RIS. (21d) implies that the RIS is deployed on the facade of buildings. (21e) indicates the decoding order constraint of NOMA technology. (21f) qualifies the transmit power constraint of the BS. We aim for modifying the radio waves based on the subsequent response from the environment. Algorithms that rely on statistical models may fail to generalize to all environment since the local topology can significantly differ from statistical prediction. To overcome the uncertainty of the environment, the RIS is supposed to learn by interacting with the environment and adapt the phase shifting policy based on the feedback of the MUs. Since the data demand of MUs are varying at each timeslot, the goal of deploying and designing the RIS is for maximizing the long-term benefits, which falls into the field of deep reinforcement learning algorithm for the reason that this algorithm can incorporate farsighted system evolution instead of myopically optimizing current benefits.

The core idea of the deep reinforcement learning approach adopted in this paper is that the designing parameters are treated as random variables, which naturally gives some joint probability distribution conditioned on MUs' date demand and mobility. Thus, this is naturally a highly dynamic scenario, which is non-trivial for conventional optimization algorithms. ML is considered to be a strong AI paradigm that can be used to empower agents by interacting with the environment and by learning from their mistakes. More explicitly, by exploiting the learning capability (learning from the environment, learning from the feedback of MUs, learning from its mistakes) of ML model, the aforementioned challenges encountered may be mitigated, leading to improved network performance. As it is non-trivial to pose the formulated problem as a supervised learning problem due to strong interactions with the environment including BSs and MUs. The RL model is capable of monitoring the reward resulting from its actions, thus it is chosen for solving the design problem in the RIS-enhanced wireless networks.

\vspace{-0.3cm}
\section{Proposed Solutions}

In this section, the LSTM-based ESN algorithm is proposed for predicting the data demand of each MU, while the ${{\text{D}}^{\text{3}}}{\text{QN}}$ algorithm is proposed for jointly deploying and designing the RIS.

\vspace{-0.3cm}
\subsection{LSTM-based ESN Algorithm for Predicting the Data Traffic Density}

In this subsection, we adopt the Recurrent Neural Network (RNN) algorithm for predicting the data traffic density based on a real dataset. A novel LSTM based ESN algorithm will be invoked, in which an RNN model is formed based on the architecture of an ESN model while leveraging LSTM units as hidden neurons.

The ESN model consists of three layers, namely input layer, hidden layer and output layer, The input weights $W_{in}$ and output weights $W_{out}$ denotes the connections between these three layers. In the ESN model, only the output weights $W_{out}$ will be adapted while the remain weights, including input weights $W_{in}$, reservoir and feedback, are randomly generated and will be fixed once the whole network is established.

In the neuron reservoir which is a sparse network in the ESN model, the typical update equations can be expressed as
\begin{align}\label{function2}
\begin{gathered}
  \tilde x(n) = tanh({W_{in}}[0:u(n)] + W \cdot x(n - 1)), \hfill \\
  x(n) = (1 - \alpha )x(n - 1) + \alpha \tilde x(n), \hfill \\
\end{gathered}
\end{align}
where $x(n)\in\mathbb{R}^{N_x}$ represents the updated version of the variable $\tilde{x}(n)$, $N_x$ denotes the the neuron reservoir size, $\alpha$ indicates the leakage rate, $tanh(\cdot)$ calculates the activation function of neurons in the reservoir, while $W \in \mathbb{R}^{N_x \cdot N_x}$ is the recurrent weight matrices.

The core advantage of the conventional ESN model is that it does not need to train the recurrent weights. However, in the conventional ESN model, the neuron reservoir consists of sparsely connected neurons, which have only a short-term memory of the previous states encountered. By invoking the LSTM units as the hidden neurons, the performance of the conventional ESN is improved with a tradeoff of increasing the computational complexity. After data echoes in the pool, it flows to the output layer, which is given by $y(n) = W_{{\text{out}}}[0;x(n)]$, where $y(n)\in\mathbb{R}^{N_y}$ denotes the network outputs and $W_{{\text{out}}}\in\mathbb{R}^{N_y\cdot(1+N_u+N_x)}$ represents the weight matrix of outputs.

\begin{remark}\label{remark:tradeofflstmesn}
By leveraging LSTM units as hidden neurons in the conventional ESN model, the LSTM-based ESN model is capable of attaining a tradeoff between the prediction performance and the computational complexity.
\end{remark}

In term of the real dataset for predicting the cellular traffic demand, the Irish Central Statistics Office periodically releases a set of demographic and socio-economic data, which are publicly available~\cite{Francesco2018Assembling}. In order to perform the ML-based prediction of data demand, the base station can exploit a dataset of the cellular traffic history. The dataset can be represented by a vector $\mathbb{U} = \left[ {u(n)\left| {\forall n \in \mathbb{N}} \right.} \right]$, where $\mathbb{N} = \left\{ {T_r,2T_r, \cdots NT_r} \right\}$ is a discrete time set of the past, and $q(n)$ is the amount of downlink data service during a time interval from $nT_r$ to $(n+1)T_r$.


\vspace{-0.5cm}

\subsection{${{\text{D}}^{\text{3}}}{\text{QN}}$ Based Algorithm for Jointly Deploying and Designing the RIS}

In this subsection, a ${{\text{D}}^{\text{3}}}{\text{QN}}$-based algorithm is proposed for determining the position and the phase shift of the RIS, as well as the power allocation from the BS to the MUs. Additionally, the DDQN-based algorithm and the DQN-based algorithm, as well as the conventional Q-learning algorithm are also discussed as benchmarks.

\subsubsection{DQN Based Algorithm for the control of the RIS}

\begin{algorithm}[!t]
\caption{Deep Q-network based algorithm for the RIS}
\label{dq1}
\begin{algorithmic}
\REQUIRE ~~\\
Q-network structure, environment simulator, replay memory $D$, minibatch size $n$.\\

\STATE \textbf{Initialize} the replay memory $D$, Q-network weights $\theta $, \\
 weights of the target network ${\theta ^*} = \theta $, and $Q(s,a)$.
\STATE \textbf{  } Randomly choose a position, phase shift and power allocation factor for the RIS.
\REPEAT
\STATE for each step of episode:\\
\STATE  \textbf{  }\textbf{  }\textbf{  }\textbf{  }The RIS chooses $a_t$ uniformly from $A$ with probability of $\varepsilon $, while chooses $a_t$ such that ${Q_\theta }({s_t},{a_t}) = {\max _{a \in A}}{Q_\theta }({s_t},{a_t})$ with probability of $(1-\varepsilon) $.
\STATE  \textbf{  }\textbf{  }\textbf{  }\textbf{  }The AV carries out action $a_t$, and observes reward $r_t$,
\STATE  \textbf{  }\textbf{  }\textbf{  }\textbf{  }The model updates state ${s_{t+1}}$;
\STATE  \textbf{  }\textbf{  }\textbf{  }\textbf{  }According to the projection hybrid NOMA precoding method, calculating the precoding metric based on current state.
\STATE  \textbf{  }\textbf{  }\textbf{  }\textbf{  }Determine the decoding order of each cluster based on the current state.
\STATE  \textbf{  }\textbf{  }\textbf{  }\textbf{  }Store transition $(s_t,a_t,r_t,{s_{t + 1}})$ and sample random minibatch of
\STATE transitions ${(s_i,a_i,r_i,{{s'}_i})}_{i \in n}$ from $D$;
\STATE  \textbf{  }\textbf{  }\textbf{  }\textbf{  }For each $i \in I$, we can obtain
\STATE  \textbf{  }\textbf{  }\textbf{  }\textbf{  }\textbf{  }\textbf{  }\textbf{  }\textbf{  }\textbf{  }\textbf{  }\textbf{  }${y_i} = {r_i} + \gamma  \cdot {\max _{a \in A}}{Q_{{\theta ^*}}}({{s'}_i},a)$;
\STATE  \textbf{  }\textbf{  }\textbf{  }\textbf{  }Perform a gradient descent step
\STATE  \textbf{  }\textbf{  }\textbf{  }\textbf{  }$\theta  \leftarrow \theta  - {a_t} \cdot \frac{1}{I}\sum\limits_{i \in n} {[{y_i} - {Q_\theta }({s_i},{a_i})] \cdot {\nabla _\theta }{Q_\theta }({s_i},{a_i})} $;
\STATE  \textbf{  }\textbf{  }\textbf{  }\textbf{  }$\theta \leftarrow {\theta ^*}$.
\STATE  \textbf{end}
\UNTIL{$s$ is terminal}
\ENSURE Action-value function ${Q_\theta }$ and policy $J$.
\end{algorithmic}
\end{algorithm}

In this subsection, the concept of the DQN-based algorithm for the deployment and control of the RIS will be first discussed. In the DQN-based model, the BS acts as an agent. Since a controller is installed, the BS can control both the power allocation policy from the BS to MUs and the RIS's position and phase shift. At each time slot $t$, the BS periodically observes the state $s_t$ of the RIS-enhanced system, from the state space, $S$. The state-space consists of the phase shift of the RIS, the allocated power to each MU, as well as coordinates of both the RIS and MUs. Accordingly, the BS carries out an action, $a_t$, from the action space, $A$, selecting the optimal choice based on policy, $J$. The action space consists of changing positions and varying phase shifts of the RIS, as well as varying the allocated power. The decision policy $J$ in the DQN model is determined by a Q-function, $Q(s_t,a_t)$. The principle of the policy is for carrying out an action that makes the DQN model obtain the maximum Q-value at each time slot. Following the chosen action, the state of the DQN model transmits to a new state $s_{t+1}$ and correspondingly the agent receives a penalty/reward, $r_t$, determined by the formulated objective function. Thus, maximizing the cumulative reward is equivalent to maximizing the average energy efficiency.

During the process of learning, the state-action value function for the agent can be iteratively updated, which is calculated as
\begin{align}\label{Pout_n}
{\begin{gathered}
  {Q_{t + 1}}({s_t},{a_t}) \leftarrow (1 - \alpha ) \cdot {Q_t}({s_t},{a_t}) +\alpha  \cdot \left[ {{r_t} + \gamma  \cdot {{\max }_a}{Q_t}({s_{t + 1}},a)} \right] \hfill \\
\end{gathered} },
\end{align}
where $\alpha $ represents the learning rate while $\gamma $ denotes the discount factor.

The learning process is divided into episodes, and at each time slot, the BS is supposed to figure out the optimal action in terms of the energy efficiency of the system. This optimal value function is the solution to the following set of equations.
\begin{align}\label{Q*}
{ {{Q}^{*}}(s,a)={{\mathbb{E}}_{{{s}'}}}\left[ r+\gamma ma{{x}_{{{a}'}}}{{Q}^{*}}({s}',{a}')\left| s,a \right. \right] },
\end{align}
where ${Q}^{*}(s,a)$ is the desired value function such that $Q\to {{Q}^{*}}$.

The DRL model can be considered as the "deep" version of conventional RL model. The DQN algorithm improves the conventional Q-learning algorithm by invoking deep neural networks as the approximator for the Q-learning model. The Q-table is approximated by a neural network (or multiple neural networks) with weights ${\rm{\{ }}\theta {\rm{\} }}$ as a Q-function, while Q-values $Q(s_t,a_t)$ is the outputs of the neural network. The value of $\theta $ is updated at each timeslot by minimizing the defined loss function
\begin{align}\label{Qfunction}
{{\rm{Loss}}(\theta ) = {\sum {\left( {y - Q\left( {{s_t},{a_t},\theta } \right)} \right)} ^2}},
\end{align}
where $y = {r_t} + \gamma  \cdot \mathop {\max }\limits_{a \in A} {Q_{{\rm{old}}}}\left( {{s_t},{a_t},\theta } \right)$.

\begin{figure} [t!]
\centering
\includegraphics[width=2.8in]{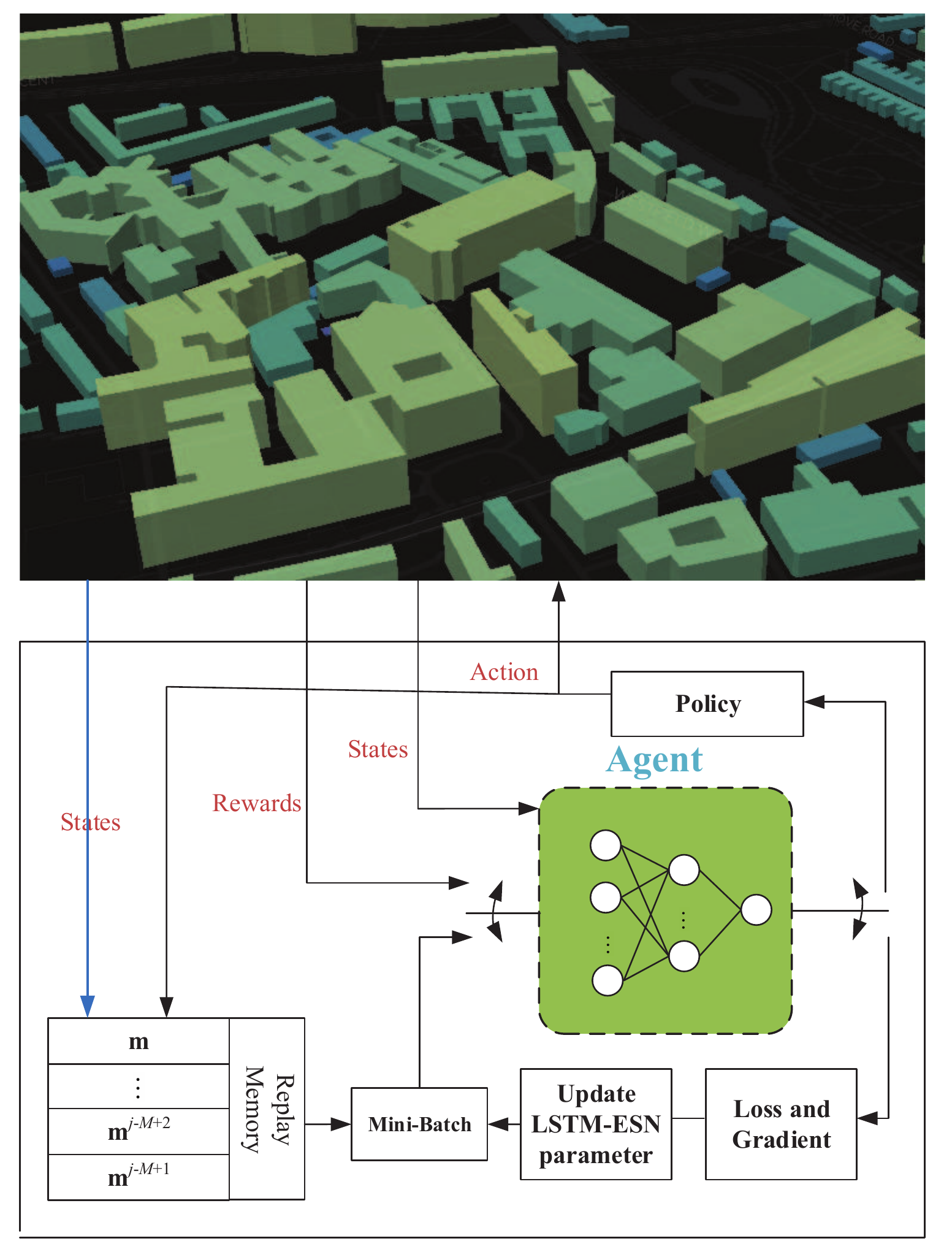}
\caption{Illustration of the ${{\text{D}}^{\text{3}}}{\text{QN}}$ based algorithm.}\label{ddqn}
\end{figure}

\subsubsection{DDQN Based Algorithm for the control of the RIS}

The key feature of the DQN algorithm is that the DQN model can incorporate farsighted system evolution (long-term benefits) instead of focusing on current states. However, one of the limitations of the DQN model is that this model may suffer from the overestimation of action values. By leveraging the same network weight $\theta $ for both action selection and $Q$ value approximation, a nonzero lower limit of ${{\max }_{a}}Q(s,a)$ is suffered, which is denoted as~\cite{van2016deep}
 \begin{align}\label{limitmax}
{{{\max }_{a}}Q(s,a)\ge {{V}^{*}}(s)+\sqrt{\frac{C}{m-1}}},
\end{align}
where ${V}^{*}$ represents a state in which all the true optimal action values are equal at ${{V}^{*}}(s)={{Q}^{*}}(s,a)$; $C$ is the variance of the state value and $m$ denotes the number of optional actions at state s. The overestimation of ${{\max }_{a}}Q(s,a)$ results in errors during the process of network training, which eventually affects the learning performance of the Q-network model. Therefore, the idea of double Q-learning is applied~\cite{van2016deep}, and a DDQN-based algorithm with the concept of double Q-learning is discussed for solving the deployment and control problem of the RIS.

In the DDQN model, the primary network is invoked to chose an action, and the target network is also adopted to generate the target Q-value for that action, instead of taking the max-over Q-values when computing the target-Q value. Therefore, the equation of Q-target can be written as
\begin{align}\label{tq}
{{y_i} = {r_i} + \gamma  \cdot {Q_{{\theta ^*}}}\left( {{{s'}_i},\arg \max \left( {Q({{s'}_i},{{a'}_i};\theta _i^{})} \right)} \right)}.
\end{align}

Two neural network models of the same architecture are established: Q estimation network and the Q target network. The network parameters are randomly initialized with normal distribution, of which the mean value is 0. A target network is used to generate the target-Q values that will be used to compute the loss for every action during training. Although not fully decoupled, the target network in the DQN architecture provides a natural candidate for the second value function, without having to introduce additional networks.

\begin{algorithm}[!t]
\caption{${{\text{D}}^{\text{3}}}{\text{QN}}$ based algorithm for the RIS}
\label{dqalgorithm}
\begin{algorithmic}
\REQUIRE ~~Q-network structure, environment simulator, \\
\textbf{  }\textbf{  }\textbf{  }\textbf{  }\textbf{  }\textbf{  }\textbf{  }\textbf{  }\textbf{  }\textbf{  }\textbf{  }replay memory $D$, minibatch size $n$.\\

\STATE \textbf{Initialize} the replay memory $D$, Q-network weights $\theta $, \\
\textbf{  }\textbf{  }\textbf{  }\textbf{  }\textbf{  }\textbf{  }\textbf{  }\textbf{  } \textbf{  }\textbf{  }\textbf{  }\textbf{  }weights of the target network ${\theta ^*} = \theta $, and $Q(s,a)$.
\STATE Randomly choose a position, phase shift and power allocation factor for the RIS.
\REPEAT
\STATE for each step of episode:\\
\STATE  \textbf{  }\textbf{  }\textbf{  }\textbf{  }The RIS chooses $a_t$ uniformly from $A$ with probability of $\varepsilon $, while chooses $a_t$ such that ${Q_\theta }({s_t},{a_t}) = {\max _{a \in A}}{Q_\theta }({s_t},{a_t})$ with probability of $1-\varepsilon $.
\STATE  \textbf{  }\textbf{  }\textbf{  }\textbf{  }Execute action $a_t$, observe reward $r_t$, update state ${s_{t+1}}$;
\STATE  \textbf{  }\textbf{  }\textbf{  }\textbf{  }According to the projection hybrid NOMA precoding method, calculating the precoding metric based on current state.
\STATE  \textbf{  }\textbf{  }\textbf{  }\textbf{  }Determine the decoding order of each cluster based on the current state.
\STATE  \textbf{  }\textbf{  }\textbf{  }\textbf{  }Store transition $(s_t,a_t,r_t,{s_{t + 1}})$ in $D$;
\STATE  \textbf{  }\textbf{  }\textbf{  }\textbf{  }Experience replay: Sample random minibatch of
\STATE  \textbf{  }\textbf{  }\textbf{  }\textbf{  }transitions ${(s_i,a_i,r_i,{{s'}_i})}_{i \in n}$ from $D$;
\STATE  \textbf{  }\textbf{  }\textbf{  }\textbf{  }For each $i \in n$, compute the target
\STATE  \textbf{  }\textbf{  }\textbf{  }\textbf{  }\textbf{  }\textbf{  }\textbf{  }\textbf{  }\textbf{  }\textbf{  }\textbf{  }${y_i} = {r_i} + \gamma  \cdot {Q_{{\theta ^*}}}\left( {{{s'}_i},\arg \max \left( {Q({{s'}_i},{{a'}_i};\theta _i^{})} \right)} \right)$;
\STATE  \textbf{  }\textbf{  }\textbf{  }\textbf{  }Update Q-network: Perform a gradient descent step
\STATE  \textbf{  }\textbf{  }\textbf{  }\textbf{  }$\theta  \leftarrow \theta  - {a_t} \cdot \frac{1}{n}\sum\limits_{i \in n} {[{y_i} - {Q_\theta }({s_i},{a_i})] \cdot {\nabla _\theta }{Q_\theta }({s_i},{a_i})} $;
\STATE  \textbf{  }\textbf{  }\textbf{  }\textbf{  } \textbf{if} ${m \le c}$
\STATE  \textbf{  }\textbf{  }\textbf{  }\textbf{  }\textbf{  }\textbf{  } \textbf{  }\textbf{  } $\varepsilon \left( m \right){\text{ = }}a\left( {\cos \left( {\frac{m}{{2c}}\pi } \right)} \right) + b$;
\STATE  \textbf{  }\textbf{  }\textbf{  }\textbf{  } \textbf{else if} ${m > c}$
\STATE  \textbf{  }\textbf{  }\textbf{  }\textbf{  }\textbf{  }\textbf{  }\textbf{  }\textbf{  } $\varepsilon \left( m \right){\text{ = }}0$;
\STATE  \textbf{  }\textbf{  }\textbf{  }\textbf{  } \textbf{end if}
\STATE  \textbf{end}
\STATE  Define policy $J$ as the greedy policy with respect to ${Q_\theta }$.
\UNTIL{$s$ is terminal}
\ENSURE Action-value function ${Q_\theta }$ and policy $J$.
\end{algorithmic}
\end{algorithm}

\subsubsection{${{\text{D}}^{\text{3}}}{\text{QN}}$ Based Algorithm for Designing the RIS}

The performance of the DRL algorithm is influenced by striking a tradeoff between exploration and exploitation. Exploration indicates that the agent can select an action (not the calculated optimal action) with non-zero probability in each encountered state for learning the environment, while exploitation, on the other hand, is aiming for employing the current knowledge of the agent to achieve relatively good performance by selecting greedy actions. In an effort to attain a tradeoff between exploration and exploitation, several strategies have been proposed, which range from full exploration to full exploitation in differing quantities, namely, random policy, greedy policy, $\epsilon $-greedy policy, decaying $\epsilon $-greedy policy, and softmax policy.

One efficient method to balance exploration and exploitation in DQN is $\epsilon $-greedy exploration. The deployment and control policy in the proposed DDQN model is chosen according to the $\epsilon $-greedy policy. More specifically, at each timeslot, the action that maximizes the Q-value of the proposed DDQN model is carried out with a high probability of $1-\epsilon $, while the other policies are chosen with a low probability for avoiding being stuck in a local maximum, i.e.,

\begin{align}\label{sq7}
{\begin{gathered}
  Pr(J=\widehat{J}) =\left\{\begin{matrix}
1-\epsilon,  & \widehat{a}=\text{argmax}Q\left ( s,a \right ),
\\
\epsilon /\left ( \left | a \right |-1 \right ),& otherwise.
\end{matrix}\right.
\end{gathered}}
\end{align}

The strategy invoked in the proposed ${{\text{D}}^{\text{3}}}{\text{QN}}$ model is also based on the $\epsilon $-greedy strategy. However, the $\epsilon $ value decays over time instead of being constant, which indicates that the strategy starts with a high $\epsilon $ value, and thus a high exploration rate. Over time this $\epsilon $ grows ever smaller until it fades, optimally as the policy has converged so that an optimal policy can be executed without having to take further (possibly suboptimal) exploratory actions.

\begin{remark}\label{remark:anye}
For any $\varepsilon$ there exists a time $T_\varepsilon$ such that for any $t>T_\varepsilon$ we have that the policy $J$ defined by incremental Q-learning model is $\varepsilon$-optimal with probability 1.
\end{remark}

In the decaying $\epsilon $-greedy strategy, we leverage the function
 \begin{align}\label{epsilonvalue}
{\varepsilon \left( m \right) = \left\{ {\begin{array}{*{20}{c}}
  {a\left( {\cos \left( {\frac{m}{{2c}}\pi } \right)} \right) + b,}&{m \le c,} \\
  0,&{m > c,}
\end{array}} \right.}
\end{align}
for $\epsilon $, where $c$ represents a number of matches, which is invoked for controlling the decaying speed of $\epsilon $, $m$ denotes the current match count, while $a$ and $b$ are with constant value for deciding the range of $\epsilon $, and $\varepsilon  \in \left[ {b,{\kern 1pt} {\kern 1pt} {\kern 1pt} a + b} \right]$, $a,b \ge 0$ and $a+b \le 1$.

In contrast to the conventional $\epsilon $-greedy policy based DQN model, the value of $\epsilon $ can be automatic decided while the actions carried out by the ${{\text{D}}^{\text{3}}}{\text{QN}}$ model is capable of converging to the optimal ones eventually.

\vspace{-0.5cm}

\subsection{State-Action Construction of the ${{\text{D}}^{\text{3}}}{\text{QN}}$ model}

Before invoking the ${{\text{D}}^{\text{3}}}{\text{QN}}$ algorithm for solving the proposed problem, it needs to represent specifically the states, action and the reward function in the proposed RIS-enhanced networking framework.

\subsubsection{State in the ${{\text{D}}^{\text{3}}}{\text{QN}}$ model}

First, we define the state of the RIS-enhanced system. State $s_t$ (at decision epoch $t$) consists of four parts: 1) ${\theta _{l,n}}(t) \in \left[ {0,2\pi } \right],{\kern 1pt}{\kern 1pt}{\kern 1pt}n \in \mathcal{N}$, the current phase shift of each reflecting elements in the RIS; 2) $c_l^I(t)=[x_l^I(t),y_l^I(t),z_l^I(t)]^T$, the current 3D position of the RIS; 3) ${c_k^U}(t) = {[{x_k^U}(t),{y_k^U}(t)]^T},{\kern 1pt}{\kern 1pt}{\kern 1pt}k \in \mathcal{K}$, the current 2D position of each MUs; 4) $p_k(t),{\kern 1pt}{\kern 1pt}{\kern 1pt}k \in \mathcal{K}$, the current power allocated from the BS to each MUs. Formally, the state $s_t=[\theta _{l,1}(t), \cdots, \theta _{l,N}(t); c_l^I(t); {c_1^U}(t) , \cdots,$ $ {c_K^U}(t); p_1(t),\cdots,p_K(t) ]$, which has a cardinality of $(N+2K+3)$.

\subsubsection{Action in the ${{\text{D}}^{\text{3}}}{\text{QN}}$ model}

In terms of actions in the ${{\text{D}}^{\text{3}}}{\text{QN}}$ model, the main operations of the RIS-enhanced system are to assign position and phase shift varies to the RIS, and to allocate power to the MUs, so that the data rate for users is improved. Action  $a_t$ (at decision epoch $t$) consists of three parts: 1) $\Delta {\theta _{l,n}}(t)  \in  \left\{ { - \frac{\pi }{{10}},{\kern 1pt} {\kern 1pt} {\kern 1pt} 0,{\kern 1pt} {\kern 1pt} {\kern 1pt} \frac{\pi }{{10}}} \right\}$, the variable quantity of the $n$-th reflecting element's phase shift; 2) $\Delta c_l^I(t) \in \left\{ {( - 1,0,0),{\kern 1pt} {\kern 1pt} {\kern 1pt} (1,0,0),{\kern 1pt} {\kern 1pt} {\kern 1pt} (0,0,0),{\kern 1pt} {\kern 1pt} {\kern 1pt} (0, - 1,0),{\kern 1pt} {\kern 1pt} {\kern 1pt} (0,1,0)} \right\}$, the moving direction and distance for RIS $l$. Explicitly, (1,0,0) means that the RIS moves right in the simulated grid world; (-1,0,0) indicates that the RIS moves left; (0,1,0) represents that the RIS moves forward; (0,-1,0) denotes that the RIS moves backward; (0,0,0) implies that the RIS keep static; 3) $\Delta {p_k}(t) \in \left\{ { - \widetilde p,0,\widetilde p} \right\}$, the variable quantity of the $k$-th MU's transmit power. Formally, the action $a_t=[\Delta {\theta _{l,1}}(t), \cdots, \Delta {\theta _{l,N}}(t); \Delta w_l^I(t); \Delta {p_1}(t),\cdots, \Delta {p_K}(t)]$, which has a cardinality of $(3N+3K+5)$.

\begin{remark}\label{remark:rdecoding}
At each timeslot, the received SINR of each MU has to be calculated for determining the dynamic decoding order before updating the state in the ${{\text{D}}^{\text{3}}}{\text{QN}}$ model.
\end{remark}

\subsubsection{Reward function in the ${{\text{D}}^{\text{3}}}{\text{QN}}$ model}

The beneficial design of the reward/penalty function is directly related to the energy efficiency of the system. When the BS carries out an action that improves energy efficiency, it receives a positive reward. By taking any other actions, which lead to a reduction in energy efficiency, the BS receives a penalty. Thus, the reward function is expressed as
\begin{align}\label{reward}
{\begin{gathered}
  r(t){\text{   =   }}\Delta {\eta _{EE}}(t) = \frac{{\sum\limits_{l = 1}^L {\left[ {{Q_{l,a}}\left( t \right) + {Q_{l,b}}\left( t \right)} \right]} }}{{\sum\limits_{l = 1}^L {\left( {{{\left\| {{{\boldsymbol{w}}_{l,a}}\left( t \right)} \right\|}^2} + {{\left\| {{{\boldsymbol{w}}_{l,b}}\left( t \right)} \right\|}^2}} \right)}  + K{P_{{\text{MU}}}}\left( t \right) + {P_{{\text{BS}}}}\left( t \right) + N{P_n}\left( t \right)}} \hfill \\
   - \frac{{\sum\limits_{l = 1}^L {\left[ {{Q_{l,a}}\left( {t - 1} \right) + {Q_{l,b}}\left( {t - 1} \right)} \right]} }}{{\sum\limits_{l = 1}^L {\left( {{{\left\| {{{\boldsymbol{w}}_{l,a}}\left( {t - 1} \right)} \right\|}^2} + {{\left\| {{{\boldsymbol{w}}_{l,b}}\left( {t - 1} \right)} \right\|}^2}} \right)}  + K{P_{{\text{MU}}}}\left( {t - 1} \right) + {P_{{\text{BS}}}}\left( {t - 1} \right) + N{P_n}\left( {t - 1} \right)}} .\hfill \\
\end{gathered} }
\end{align}

It can be observed from \eqref{reward} that, maximizing the long-term sum rewards tends to maximize the long-term energy efficiency of the RIS-enhanced cellular networks.

\subsection{Analysis of the Proposed Algorithm}

\subsubsection{Convergence of the proposed algorithm}

Four steps are taken for analyzing the convergence of the ${{\text{D}}^{\text{3}}}{\text{QN}}$ algorithm. The first step is proving that the general Q-learning approach is indeed capable of converging to the optimal state; the second step is proving that the double Q-learning algorithm with decaying exploration learning policy is capable of converging to the optimal state; the third step is proving that the double Q-learning algorithm with persistent exploration learning policy is capable of converging to the optimal state; the fourth step is showing that the neural network approach succeeds in identifying the nonlinear Q-values generated by the Q-learning iteration.

\begin{proposition}\label{Qproof}
The decaying double deep Q-network approach is capable of converging to the optimal Q value.
\end{proposition}

Since ${{\Delta _t}}$ converges to zero under the assumptions stipulated in~\cite{melo2001convergence}, the conventional Q-learning model converges to the optimal Q-function as long as $0 \le {\alpha _t} \le 1,{\kern 1pt} {\kern 1pt} {\kern 1pt} {\kern 1pt} \sum\limits_t {{\alpha _t} = \infty {\kern 1pt} } $ and $\sum\limits_t {\alpha _t^2 < \infty {\kern 1pt} } $. According to Lemma 2 in~\cite{hasselt2010double}, the double Q-learning model can converge to an optimal state. According to Corollary 5.4 in~\cite{even2002convergence}, for any $\varepsilon$ there exists a time $T_\varepsilon$ such that for any $t>T_\varepsilon$ we have that the policy $J$ defined by incremental Q-learning model is $\varepsilon$-optimal with probability 1, which indicates that the double Q-learning algorithm with persistent exploration learning policy is capable of converging to the optimal state. By following the Stone-Weierstrass Theorem~\cite{timofte2018stone}, it can be observed that if the neural network is large enough and the initial conditions are appropriately chosen, the neural network is capable of approximating any non-linear continuous function. Overall, the decaying double deep Q-network approach is capable of converging to the optimal Q value.

\subsubsection{Complexity of the proposed algorithm}

The complexity of the proposed ${{\text{D}}^{\text{3}}}{\text{QN}}$ algorithm is mainly related to the CNN model and the learning process. As demonstrated in~\cite{he2015convolutional}, the CNN's computational complexity is $O(\Xi )=O\left[ {{f}_{1}}\left( n_{2}^{2}{{\left( {{n}_{1}}-{{n}_{2}}+1 \right)}^{2}}+{{f}_{2}}n_{3}^{2}{{\left( {{n}_{1}}-{{n}_{2}}-{{n}_{3}}+2 \right)}^{2}} \right) \right]$, where $i$ represents the Conv layer number. For the first layer, $f_1$ filters are needed, while each filter has a size of $({{n}_{2}}\times {{n}_{2}})$, and outputs a ${{f}_{1}}{{\left( {{n}_{1}}-{{n}_{2}}+1 \right)}^{2}}$ feature map. Similarly, the second Conv layer has $f_1$ filters each of size $({{n}_{3}}\times {{n}_{3}})$, while each filter inputs a $({{n}_{2}}\times {{n}_{2}})$ element matrix and outputs a ${{f}_{2}}{{\left( {{n}_{1}}-{{n}_{2}}-{{n}_{3}}+2 \right)}^{2}}$ feature map. In terms of the learning process, the computational complexity is related to the number of timeslots $T$ and episodes $N_e$. Thus, the total computational complexity of the ${{\text{D}}^{\text{3}}}{\text{QN}}$ algorithm is $O(TN_e\Xi)$.

\begin{figure} [t!]
\centering
\includegraphics[width=3in]{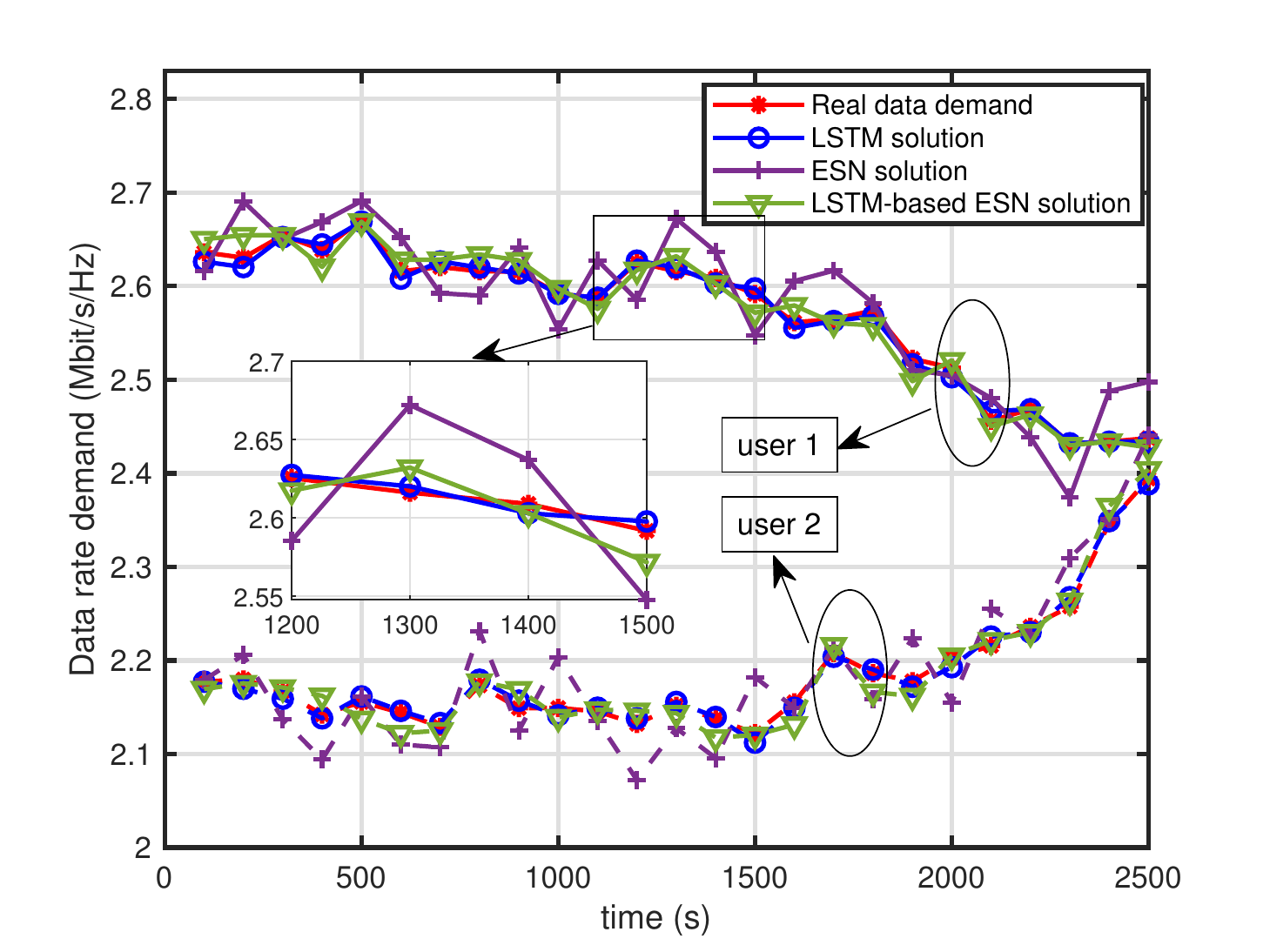}
\caption{Prediction of users' data demand over different algorithms.}\label{LSTMESNpre}
\end{figure}

\vspace{-0.5cm}
\section{Numerical Results}

In this section, we verify the efficiency of the proposed ${{\text{D}}^{\text{3}}}{\text{QN}}$ algorithm, as well as the performance of the NOMA-RIS enhanced wireless system. In the simulations, the MUs are randomly located, while the BS is deployed at the center of the square region with a side length of 100m. The maximal transmit power of the BS is 20dBm. We consider the case that 6 MUs are partitioned into 3 clusters. The other simulation parameters are given in Table~\ref{table:parameters}. In terms of the small scale fading, we follow the simulation parameter setting of~\cite{mu2019exploiting}. On this basis, we analyze the energy efficiency of the NOMA-RIS enhanced system, data demand prediction of the MUs, the deployment and control policy of the RIS. All the results are averaged over 100 independent channel realizations.

\begin{table*}[htbp]\footnotesize
\caption{ Simulation parameters}
\centering
\begin{tabular}{|l||r||r|}
\hline
\centering
 Parameter & Description & Value\\
\hline
\centering
 $P_{\text {MU}}$ & Hardware power dissipation for MUs& 10dBm\\
\hline
\centering
 $P_{\text{BS}}$ & Hardware power dissipation for BS & 9dBW\\
\hline
\centering
 $P_n$ & Power dissipation of varactor diode & 0.25W\\
\hline
\centering
 $\alpha_{BM} $ & Path loss exponent for BS-MU link & 3.5~\cite{zhang2019capacity,mu2019exploiting}\\
  \hline
  \centering
 $\alpha_{BS} $ & Path loss exponent for BS-RIS link & 2.2~\cite{zhang2019capacity,mu2019exploiting}\\
  \hline
  \centering
 $\alpha_{SM} $ & Path loss exponent for RIS-MU link & 2.8~\cite{zhang2019capacity,mu2019exploiting}\\
  \hline
\centering
 $C_0$ & Path loss when $d_0=1$ & -30dB\\
 \hline
\centering
 $N_0$ & Noise power spectral density& -169dBm/Hz \\
 \hline
\centering
 $N_x$ & Size of neuron reservoir & 2000 \\
 \hline
 \centering
 $\alpha_l $ & Learning rate & 0.01\\
 \hline
 \centering
 $\beta $ & Discount factor & 0.7\\
 \hline
\end{tabular}
\label{table:parameters}
\end{table*}



\textit{1) Prediction of users' tele-traffic requirement:} Fig.~\ref{LSTMESNpre} characterizes the prediction accuracy of two MUs in the same cluster over different algorithms. It can be observed that the proposed LSTM-based ESN algorithm outperforms the ESN algorithm, while its prediction accuracy is slightly lower than the LSTM algorithm. However, the computational complexity of learning LSTM models is dominated by ${n_c} \times \left( {{n_c} + {n_o}} \right)$ factor with $n_c$ representing the number of memory cells and $n_o$ denoting the number of output units. When tackling predicting tasks with a large number of output units and require a large number of memory cells for storing temporal contextual information, the LSTM algorithm becomes computationally expensive. On the other hand, the proposed LSTM-based ESN algorithm is capable of achieving striking a tradeoff between the prediction accuracy and computational complexity, while guaranteeing the prediction performance for the users' tele-traffic requirement.

\begin{figure}[H]
	\begin{minipage}[t]{0.46\textwidth}
		\centering
		\includegraphics[scale=0.5]{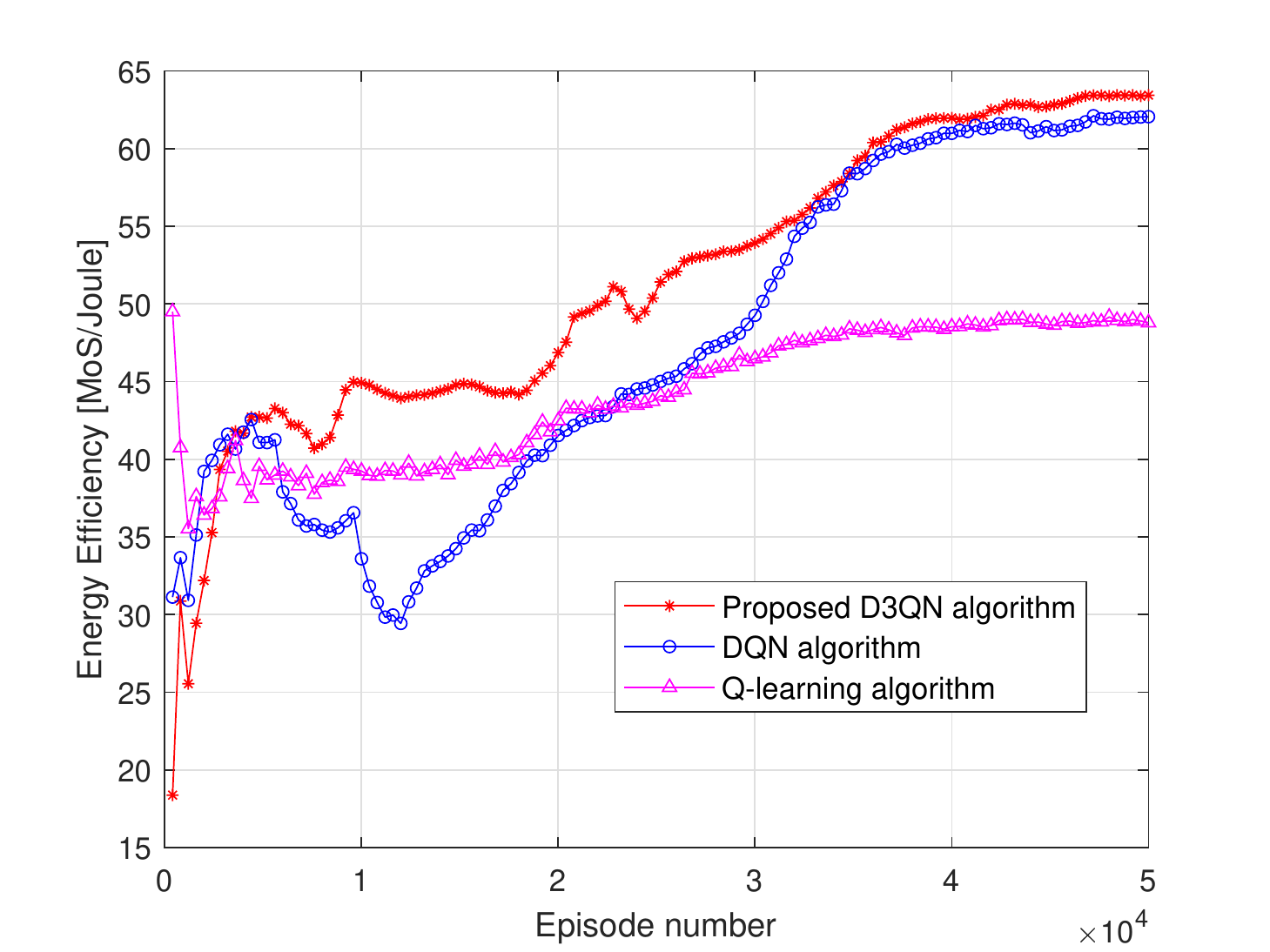}
		\caption{Instantaneous energy efficiency over episode number. (Learning rate is 0.070)\label{convergence}}
	\end{minipage}
	\qquad
	\begin{minipage}[t]{0.46\textwidth}
		\centering
		\includegraphics[scale=0.5]{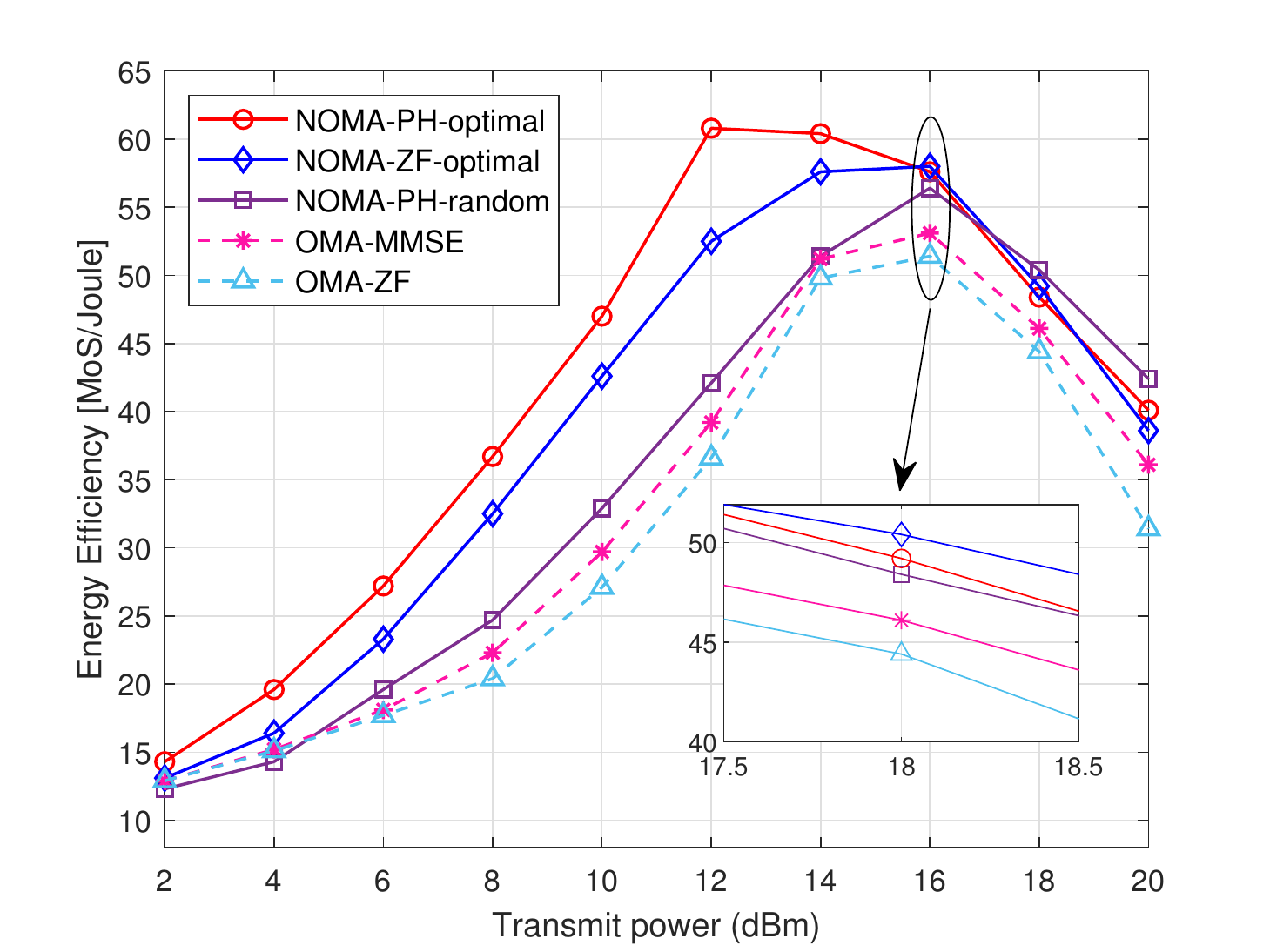}
		\caption{Energy efficiency (in MOS/Joule) over transmit power.(the number of antenna of the BS is 6, the number of reflecting elements is 16)\label{EEvsP}}
	\end{minipage}
\end{figure}

\textit{2) Convergence of the proposed algorithm:} Fig.~\ref{convergence} demonstrates the convergence of the proposed ${{\text{D}}^{\text{3}}}{\text{QN}}$ algorithm, while the DQN-based algorithm and the conventional Q-learning algorithm are also demonstrated as benchmarks. The result of the DQN algorithm is obtained by searching for an optimal $\epsilon $. It can be observed from fig.~\ref{convergence} that both the ${{\text{D}}^{\text{3}}}{\text{QN}}$-based algorithm and DQN-based algorithm are capable of converging, while the Q-learning based algorithm can not converge to the optimal state due to the huge state space and action space of the Q-table in the Q-learning model. Additionally, the proposed ${{\text{D}}^{\text{3}}}{\text{QN}}$ algorithm outperforms the DQN algorithm due to the advantage that the proposed ${{\text{D}}^{\text{3}}}{\text{QN}}$ algorithm invokes the concept of double Q-learning approach and decaying $\epsilon $-greedy policy.

\textit{3) Energy efficiency versus transmitting power:} Fig.~\ref{EEvsP} characterizes the energy efficiency of the RIS-enhanced system versus the transmit power of the BS. It can be observed that the energy efficiency of the system raises sharply, as the transmit power increases from 2dBm to 12dBm. However, since the maximal satisfaction level of MUs is 5. thus, after the transmit power is high enough for each MU, increasing the transmit power of the BS can not enhance the sum MOS of the MUs, but rises the energy consumption, which leads to the decline of the energy efficiency. It can also be observed that the PH-NOMA precoding method outperforms the ZF precoding method. The reason is that the ZF precoding method is a single beam per group approach, if the data demand of the strong user is comparable to that of the weak user, the power allocation factor $a_{l,2}$ has to be sufficiently close to 0 in terms of fairness. The NOMA-PH-random line in the fig.~\ref{EEvsP} represents that the decoding order in the NOMA-RIS enhanced system is fixed. Since the phase shift of the RIS is dynamic changed, the dynamic decoding order has to be determined for guaranteeing SIC performance successfully. Thus, the NOMA-PH-optimal approach, which re-determines the decoding order at each timeslot, outperforms the NOMA-PH-random approach with a fixed decoding order. This phenomenon is also confirmed by the insights provided in \textbf{Remark 6}. Additionally, the NOMA-assisted RIS system is capable of obtaining better performance in terms of energy efficiency compared to the OMA-assisted RIS system.

\begin{figure}[H]
	\begin{minipage}[t]{0.46\textwidth}
		\centering
		\includegraphics[scale=0.5]{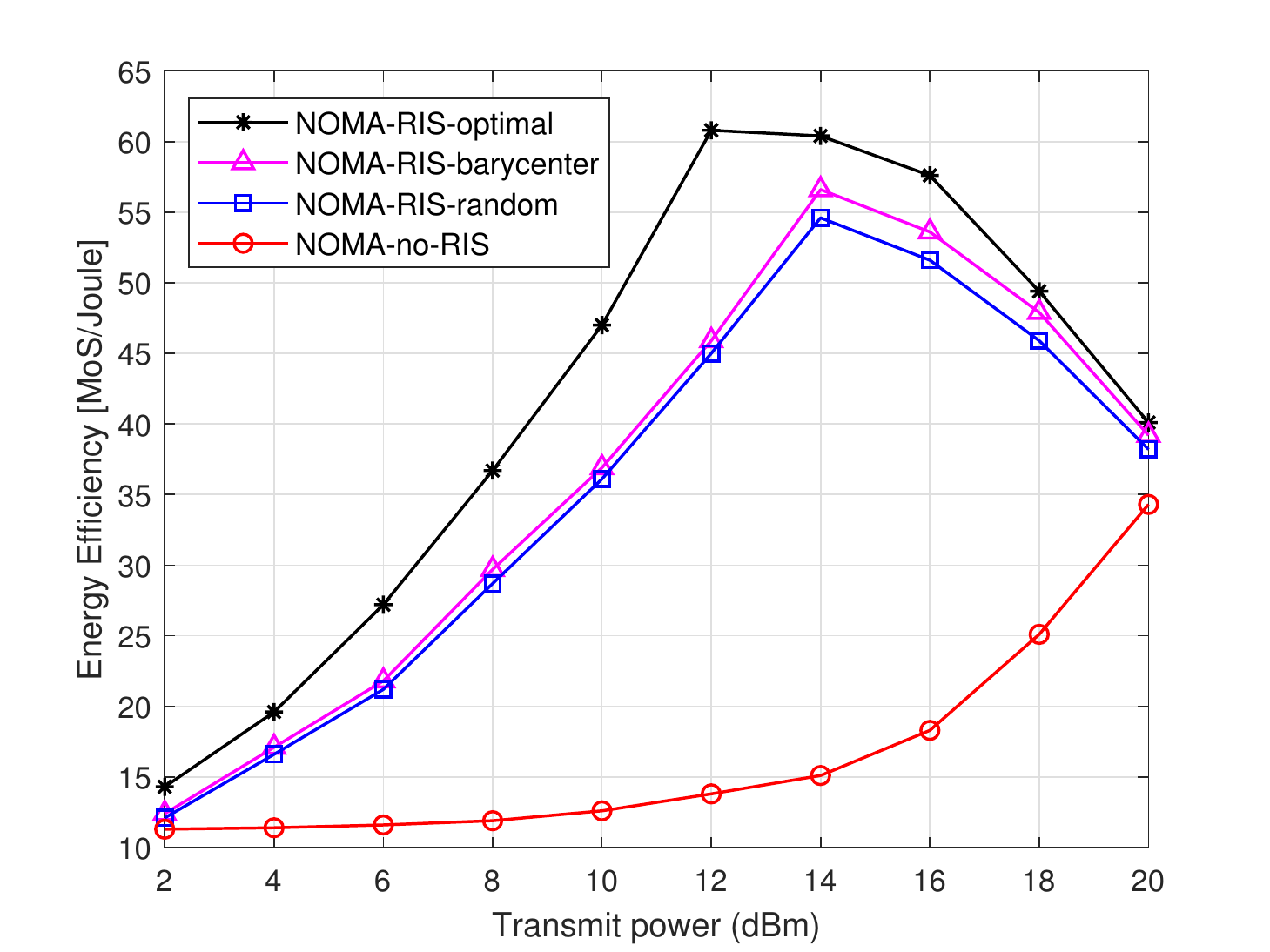}
		\caption{Energy efficiency (in MOS/Joule) with and without RIS.(the number of antenna of the BS is 6, the number of reflecting elements is 16)\label{EEvsnoIRS}}
	\end{minipage}
	\qquad
	\begin{minipage}[t]{0.46\textwidth}
		\centering
		\includegraphics[scale=0.5]{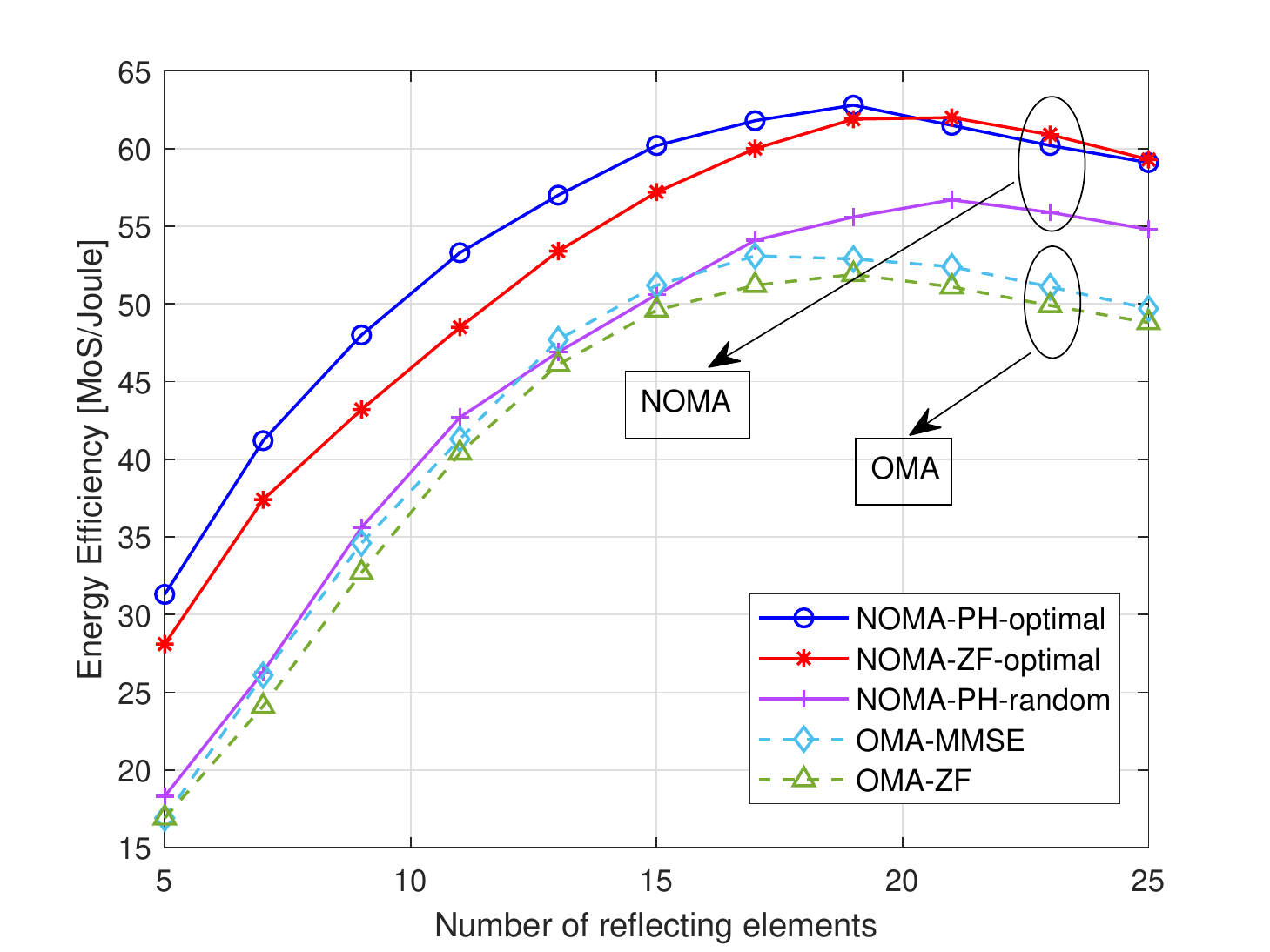}
		\caption{Energy efficiency (in MOS/Joule) vs reflecting elements number.(the number of antenna of the BS is 6, the transmit power of the BS is 16dBm)\label{EEvsN}}
	\end{minipage}
\end{figure}

\textit{4) Impact of the RIS in terms of energy efficiency:} Fig.~\ref{EEvsnoIRS} characterizes the energy efficiency of the system in both NOMA networks with the assistance of the RIS and without the RIS. It can be observed that the energy efficiency of the system is enhanced by employing the RIS. The NOMA-RIS-barycenter line denotes that the RIS is placed at the barycenter of all MUs, the NOMA-RIS-random line represents that the RIS is randomly deployed, while the NOMA-RIS-optimal line denotes that the RIS is deployed at the optimal position derived from the proposed ${{\text{D}}^{\text{3}}}{\text{QN}}$ algorithm. The results of fig.~\ref{EEvsnoIRS} confirm that there exists an optimal position for the RIS as far as the energy efficiency of the RIS-enhanced system is considered. The performance of the RIS-enhanced system is capable of being improved with deploying the RIS at the optimal position compared to random deployment and placing it at the barycenter.

\textit{5) Energy efficiency versus the number of reflecting elements:} Fig.~\ref{EEvsN} characterizes the energy efficiency of the RIS-enhanced system versus the number of reflecting elements. It can be observed that the energy efficiency of the system increases rapidly, as the number of reflecting elements increases from 5 to 18. As discussed above, for the RIS-enhanced system with a fixed transmit power, after the number of reflecting elements is enough for each user, increasing the number of reflecting elements can not enhance the sum MOS of the MUs, but rises the energy consumption due to the parameter $P_{\text{RIS}}$. The results of fig.~\ref{EEvsN} confirm that there exists an optimal number of reflecting elements for the RIS as far as the energy efficiency of the RIS-enhanced system is considered and the transmit power of the BS is fixed. This phenomenon is also confirmed by the insights provided in \textbf{Remark 3}.

\vspace{-0.5cm}
\section{Conclusions}

The joint deployment, phase shift control, power allocation and dynamic decoding order determination of the RIS-enhanced wireless system was considered with considering the particular data requirement of MUs. Additionally, NOMA technology was invoked for further enhancing the spectral efficiency. To tackle the problem formulated, an LSTM-based ESN algorithm was firstly proposed for predicting the future tele-traffic demand of MUs based on a real dataset. Secondly, a ${{\text{D}}^{\text{3}}}{\text{QN}}$-based algorithm was proposed for determining the position and control policy of the RIS. By receiving real-time feedback from the MUs and learning from its mistakes, the BS (acting as an agent in the ${{\text{D}}^{\text{3}}}{\text{QN}}$ model) was shown being capable of obtaining a policy, which determines the placement and control of the RIS for obtaining the maximal energy efficiency of the system. It was also demonstrated that the proposed ${{\text{D}}^{\text{3}}}{\text{QN}}$ algorithm was capable of converging after appropriate training. Additionally, the proposed ${{\text{D}}^{\text{3}}}{\text{QN}}$ algorithm outperformed the DQN algorithm by leveraging the concept of the double Q-learning model and decaying $\epsilon $-greedy policy. Additionally, the proposed NOMA-RIS approach outperforms the benchmarks in terms of energy efficiency.

\begin{spacing}{1.4}
\bibliographystyle{IEEEtran}
\bibliography{mybib}
\end{spacing}
\end{document}